\documentclass[aps,prd,twocolumn,nofootinbib,floatfix]{revtex4-1}
\usepackage[english]{babel}
\usepackage{setspace}
\usepackage{amsthm}
\theoremstyle{definition}

\usepackage{graphicx}
\usepackage{dcolumn}
\usepackage{multirow}
\usepackage{subfigure}
\usepackage{times,mathptm}
\usepackage{float}
\usepackage{color}
\usepackage{amsmath,amsfonts}
\usepackage{mathptmx}
\usepackage{mathrsfs}
\usepackage{bbm}
\usepackage{bm}
\usepackage{xfrac}

\newcommand{\beq}{\begin{equation}}
\newcommand{\eeq}{\end{equation}} 
\newcommand{\bea}{\begin{eqnarray}}
\newcommand{\eea}{\end{eqnarray}} 

\newcommand{\Om}{\Omega}
\newcommand{\up}{\uparrow}
\newcommand{\dn}{\downarrow}

\newcommand{\E}{\mathcal{E}}

\renewcommand{\d}{\delta}

\newcommand{\Q}{\mathbf{Q}}
\newcommand{\R}{\mathbf{R}}

\renewcommand{\r}{\rho}

\newcommand{\D}{\Delta}

\newcommand{\rhot}{\widetilde{\rho}}
\renewcommand{\th}{\theta}

\newcommand{\oh}{\frac{1}{2}}

\newcommand{\dg}{\dagger}
\newcommand{\non}{\nonumber}

\newcommand{\rf}[1]{(\ref{#1})}
\newcommand{\ra}{\rightarrow}

\usepackage{xcolor}
\definecolor{olive}{rgb}{0.3, 0.4, .1}
\definecolor{fore}{RGB}{249,242,215}
\definecolor{back}{RGB}{51,51,51}
\definecolor{title}{RGB}{255,0,90}
\definecolor{dgreen}{rgb}{0.1,0.5,0.1}
\definecolor{gold}{rgb}{1.,0.84,0.}
\definecolor{JungleGreen}{cmyk}{0.99,0,0.52,0}
\definecolor{BlueGreen}{cmyk}{0.85,0,0.33,0}
\definecolor{RawSienna}{cmyk}{0,0.72,1,0.45}
\definecolor{Magenta}{cmyk}{0,1,0,0}

\usepackage{ulem}

\begin{document}

\title{BCS states and D-wave condensates in the 2D Hubbard model}
\bigskip
\bigskip

\author{Kazue Matsuyama and Jeff Greensite}
\affiliation{Physics and Astronomy Department \\ San Francisco State
University  \\ San Francisco, CA~94132, USA}
\bigskip
\date{\today}
\vspace{60pt}
\begin{abstract}

\singlespacing

We consider states of BCS form in the 2D Hubbard model which, starting from some arbitrary point in state space in the neighborhood of a Hartree-Fock ground state, are relaxed within that BCS ansatz to local minima of the energy.  As in the Hartree-Fock approximation there are a vast number of local minima, nearly degenerate in energy.  What is new, and unlike the conventional Hartree-Fock states, is that there is a region in parameter space where these local minima are clearly associated with d-wave condensates of the form $d_{x^2-y^2}$ in the underdoped region.  There are also indications of $d_{xy}$ condensation in the overdoped region, at least in this approximation to the 2D Hubbard model, as well as condensates over a range of parameters on triangular lattices.

\end{abstract}

%
%
%
\maketitle
 
\singlespacing

\section{\label{Intro} Introduction}

  The Hubbard model is believed to hold the key to understanding important features of condensed matter physics, not least the phenomenon of high T$_\text{c}$ superconductivity in the cuprates.  Yet despite the apparent simplicity of the Hamiltonian, it is very hard to solve for even the ground state of the system (a feature which the Hubbard model shares with Yang-Mills theory).  The problem is that interesting physical phenomena are associated with regimes in which neither strong nor weak coupling expansions are trustworthy.  This has led to the development or application of a number of non-perturbative approximation methods.  The first such method, was the Hartree-Fock approximation, whose application to the Hubbard model goes all the way back to 1966 \cite{Penn}, and is very well represented in the older Hubbard model literature, cf.\ \cite{Hirsch,Poilblanc,Zaanen,Machida,Schulz1,Schulz2,Ichimura,Verges,Inui,Dasgupta,Bach,Imada,Fazekas,Xu,multiref,PRX}  for a sample.  This approximation has had a number of striking successes, such as the prediction of stripes in the cuprate phase diagram \cite{Zaanen,Machida}, but it has had limited success in applications to the overall phase diagram, particularly at finite temperatures, and the demonstration of a pairing condensate.  The approximation cannot properly account for electron correlations in this highly correlated system, and of course neglect of correlations is a feature which is built into any mean field method.   Powerful quantum Monte Carlo methods which might serve as a first principles approach are unfortunately limited to very small electron/hole dopings by the sign problem inherent in the Hubbard model \cite{Varney}.
   
    Many methods have been introduced over the years to overcome these problems.  Complex Langevin \cite{Yamamoto}, Lefshetz thimble \cite{Ulybyshev}, and density-of-state methods \cite{Korner}, have been employed against the sign problem, along with constrained path Monte Carlo methods \cite{constrainedMC}.   At present the density matrix renormalization group (DMRG) is probably the most widely used procedure \cite{DMRG1} in the field, and is now recognized to be closely related to Matrix Product States (MPS) \cite{DMRG2}.   Unfortunately, growth in computation time with lattice volume limits its application in two dimensions to strips and cylinders of modest width or circumference.   Dynamical Mean Field Theory (DMFT) treats the mean field as time dependent \cite{DMFT}, but here again spatial correlations are neglected.  There is an extension, cluster DMFT,  which attempts to include some short-range correlation, which is also the goal of the Dynamical Cluster Approximation \cite{cDMFT,DCA}, and these again incorporate some mean field elements.  This abbreviated list does not exhaust the methods which have been applied to the 2D Hubbard model.  Many of these methods remain as important tools, and new ones are still being introduced (e.g.\  Pfaffian wavefunctions \cite{Pfaffian1}).   It is not our intention here to review any of these approaches. We only point out that the number of competing procedures in current use is an indication that no single procedure has been entirely successful in treating all important aspects of the 2D Hubbard model.
    
    This perhaps justifies our introduction of yet another approximation method, intended especially for locating the pairing condensate and its symmetry in the $U/t$-density plane at zero temperature.   We are motivated in part by the traditional BCS ansatz, in which electron number is not an eigenstate of the ground state. Then there can be non-zero expectation values for pairing operators such as $c^\dg_\up(x) c^\dg_\dn(y)$, which do not commute with total electron number, where $c^\dg_s,c_s$ are the usual electron creation/destruction operators with spin $s={\up,\dn}$.  Obviously a physical system with a fixed number of electrons is not precluded from pair condensation, but the investigation requires computing the expectation value of widely separated pairing operators, which is more difficult. 

    The intention of our approximation is to merge some of the good features of both the standard BCS approach and the Hartree-Fock method.  The idea, motivated by BCS, is that only the states near the 
top of the Fermi sea are relevant for pair condensation, but we do not assume that either these states, or the states below the Fermi surface, are plane wave states.  Instead, these lower states are taken directly from a Hartree-Fock treatment.  The upper states are used to construct the BCS ansatz, and these states are allowed to mix with one another. Occupation number is varied with the restriction that the expectation value of particle number is preserved.  The idea of applying BCS to the cuprates is of course not new, and derives  from Anderson \cite{Anderson_1987}, who combined a particular form of the standard BCS ansatz with a Gutzwiller projection, and variation to an energetic minimum.  Closer to own proposal is the generalized Hartree-Fock wave function whose form was put forward by Bach et al.\ \cite{Bach},  although that article was mainly concerned with issues such as the existence of various symmetries and was not, to our knowledge, ever used to obtain any numerical results.  We comment further on these wave functions in the next section.

   What we find in our approach is a condensate with $d_{x^2-y^2}$ symmetry in a coupling-density region in the neighborhood of 
$U/t=4$ and lightly hole-doped density of electron density $f=0.8$ at zero temperature, and some indications of $d_{xy}$ condensation in the highly overdoped region. \footnote{The BCS approach has recently been advocated in the overdoped region  in \cite{Kivelson-BCS}.}  We also illustrate the position dependence of the condensate at certain points in the phase diagrams, and find some correlation with stripes in magnetization.  In the standard Hubbard treatment \cite{Verges,Inui,Xu} there are very many local energy minima with nearly equal energies, and in
\cite{Matsuyama:2022kam} it was suggested that this might be a genuine feature of the 2D Hubbard model, with a vast landscape of eigenstates nearly degenerate in energy, reminiscent of a spin glass.  The same multiplicity of local minima is found in our BCS-inspired approach.

We have also applied our approach to the study of condensates on triangular lattices, which appear to exist for a wide range of electron densities and coupling strengths.

Those are the main results of our calculation.   In section \ref{proc} below we describe the details of our approach.  In principle there are an enormous number of parameters that can be varied, even within this BCS ansatz, and we explain how we cut this down to a number which is tractable, at least numerically.  Our numerical results are contained in section III, with conclusions in section IV. 

\section{\label{proc} Procedure}          

   Rather than starting from the plane wave eigenstates of a free fermi gas, as in the Gutzwiller/RVB approach   {\cite{Anderson_1987,Gros,Kotliar,Paramekanti,Anderson_2004}}, we begin instead with the one-particle wave functions $\{\phi_i(x,s), i=1,2,..,2L^2\}$ obtained in the Hartree-Fock approximation, where, for $M$ electrons on an $L\times L$ periodic lattice,
\beq
|\Om_{HF} \rangle = \prod_{i=1}^M \left(\sum_{x_i,s_i} \phi_i(x_i,s_i) c_{s_i}^\dg(x_i)\right) |0\rangle
\eeq
is the minimal energy state obtained by the usual mean field methods (our particular implementation is described in
\cite{Matsuyama:2022kam}).  It is, as already remarked, not unique (local energy minima depend on initialization), and certainly not adequate to describe condensates, but it does get a few things right, such as the existence of stripes.  

    In the BCS description of normal superconductivity, pairing of electrons and the appearance of a gap and condensate is a feature which is happening among energy levels at the top of the Fermi surface; lower levels are unaffected.  With this
in mind we suggest  dividing our ansatz for the ground state into a product
of one particle states in lower-lying levels, which will not be varied, and a BCS-like state
\bea
|\Om\rangle &=& \left[\prod_{i=1}^{M-N} \left(\sum_{x_i,s_i} \phi_i(x_i,s_i) c_{s_i}^\dg(x_i)\right)\right] \non \\
  & & \times  \prod_{i=1}^{N} (a_i + b_i U^\dg_i D^\dg_i) |0\rangle \ ,
\label{Om}
\eea
where
\bea
U_i^\dg &=& \sum_{x_i} \sum_{s_i=\up,\dn} u_i(x_i,s_i) c^\dg_{s_i}(x_i) \non \\
 D_i^\dg &=& \sum_{x_i} \sum_{s_i=\up,\dn} d_i(x_i,s_i) c^\dg_{s_i}(x_i) \ ,
\eea
and the set $\{u_i(x,s), d_i(x,s)\}$ are
orthonormal one-particle wave functions to be described in more detail below. Generalized Hartree-Fock BCS wavefunctions of this form were first put forward by Bach et al.\ \cite{Bach}, and in that reference the quantities which appear in the wave function should be fixed by a Bogoliubov transformation to a state which minimizes $\langle H \rangle$.  But unless the required transformation is block diagonal in momentum space, as in the usual BCS approach, finding the required Bogoliubov transformation is not an easy task, and \cite{Bach} does not offer much guidance in this respect. Likewise, the authors of refs.\ [29-33] assume that the initial state, prior to projection, is  a traditional BCS state with a particular choice (in standard notation) of $v_k/u_k$ which depends on a few variational parameters, and definite symmetries are assumed from the start. Our approach differs in that our starting point is Hartree-Fock one-particle wavefunctions (although some of these get mixed near the Fermi surface), and we dispense with the Gutzwiller projection. No d-wave or other pairing symmetry is assumed from the start; certain symmetries seem to be an output of the calculation.
 In fact our approach may be regarded as a suggestion for obtaining the generalized Hartree-Fock state \rf{Om} in practice.

 The general idea is to start with
some random choice of $|\Om\rangle$ of this form, and then let the state systematically relax to an energy minimum,
while keeping the expectation value of electron number fixed.
The one-particle $u_i(x,s), d_i(x,s)$ wave functions lie in the Hilbert space spanned by the remaining
$\{\phi_i(x,s),  i= M-N+1,...,M+N\}$ one-particle wave functions, $N$ of which lie at or immediately below the Fermi level, and the other $N$ lie just above.  The $u_i, d_i$ wave functions are concentrated at up, down spins respectively, and the associated $a_i,b_i$ coefficients are subject to variation, preserving orthogonality.  $N$ is a free parameter, and the remainder of the states, involving the lower-lying levels, held fixed.   Apart from assuming the BCS form itself, and the Hartree-Fock initialization, this division into a product of fixed lower levels and a state subject to variation is our main approximation.  

\subsection{Initialization}

We compute the average spin of the $i$-th one particle wavefunction
\beq
            \overline{s}_i = \sum_x (\phi_i^2(x,\up) - \phi_i^2(x,\dn)) \ .
\label{sbar}
\eeq
Starting at the Fermi level $i=M$ and working downwards, we relabel the first $N/2$ states with 
$\overline{s_i} > 0$
as $\{u_j(x,s), j=1,2,..,N/2, s=\up, \dn\}$, and then the first $N/2$ states above the Fermi level, with
$\overline{s}_i>0$, as
$\{u_j(x,s), j=N/2+1,..,N\}$.  The index $j$ is in order of increasing one-particle (Hartree-Fock) energy.
The same procedure is applied to identify $N$ states $d_i(x,s)$ in the neighborhood of the Fermi surface with 
$\overline{s}_i<0$.  The remaining states $\phi_i$ below the Fermi level are renumbered, if necessary, so that the $\{\phi_i, i=1,2,...M-N\}$ do not coincide with any of the $u,d$ wavefunctions. In other words, 
$\{u_j(x,s),d_j(x,s)\}$ wavefunctions are initially elements of the original Hartree-Fock set, in the neighborhood of the Fermi surface.  This initial $u_j,d_j$ pairing of one particle wave functions of opposite average spin and approximately the same energy is of course motivated by the original BCS ansatz.

The original Hartree-Fock state would be given by the choice of coefficients
\beq
              \begin{array}{cc}
                  a_i=0, ~ b_i=1 & ~~~ 1\le i \le N/2 \cr
                  a_i=1, ~ b_i=0 & ~~~ N/2+1 \le i \le N \end{array} \ ,
\label{HF}
\eeq
and the energy for this choice is always computed at the start.  Our aim is to examine local minima in the neighborhood of this Hartree-Fock state, beginning with a random deviation from the Hartree-Fock state \rf{HF} near the Fermi surface, and then relaxing to a local minimum.  There are several constraints on the $a_i,b_i$ coefficients:  First, the expectation value of electron number is unchanged in the relaxation procedure, which requires that
\beq
N = \sum_{i=1}^{N} 2 b_i^2  \ ,
\label{number}
\eeq
 and secondly
 \beq
 0 \le |a_i|, |b_i| \le 1 ~~~\text{and} ~~~ a_i^2+b_i^2 = 1 \ .
  \label{cond}
 \eeq
 
    The initial state is the Hartree-Fock state \rf{HF}.  To obtain a random state somewhere in the neighborhood of this state, we select at random pairs of indices  $i\ne j$  in the range $1-N$.  Then define
  \bea
\d_1 &=& \max(-b_i^2,b_j^2-1) \non \\
\d_2 &=& \min(1-b_i^2,b_j^2) \non \\
\d &=&  \d_1 + x(\d_2-\d_1) \ ,
\label{d}
\eea 
where $x$ is a random number uniformly distributed between 0 and 1.  Then we choose new coefficients
 \bea
b'_i &=& \sqrt{b_i^2 +\d} \non \\
b'_j &=& \sqrt{b_j^2 -\d} \non \\
a'_{i} &=& s_1\sqrt{1-b^{'2}_{i}} \non \\
a'_{j} &=& s_2\sqrt{1-b^{'2}_{j}}  \ ,
\label{ab}
\eea
where $s_{1,2}$ are chosen to equal $\pm 1$ with equal probability.  With these rules, the conditions \rf{cond} are obeyed, and the expectation value of electron number is preserved.  Repeating this process for 100 randomly chosen pairs gives us a starting point with a new set, whose energy is generally far above that of the initial Hartree-Fock state.

\subsection{Relaxation}

The relaxation to a local energy minimum is iterative, proceeding for thousands of updates until the energy
expectation value 
${\E = \langle \Om|H|\Om\rangle}$, minus the energy due exclusively to states $\{\phi_i, i=1,M-N\}$ 
(see \rf{energy} below) is deemed to have converged by remaining unchanged, up to the fourth digit, after one thousand attempted updates.  Relaxation has both stochastic and deterministic elements.  The first step in each update is to choose at random, with uniform probability, two indices $i \ne j$ in the range $1$ to $N$.  This is the stochastic element, the rest is deterministic.  The next step is to choose new $a_{i,j},b_{i,j}$ coefficients for this pair from \rf{d} and \rf{ab}, except that the variable $x$ in \rf{d} is no longer a random number, but is chosen numerically to minimize the energy expectation value which depends on these coefficients.\footnote{Factors $s_1,s_2$ in \rf{ab} are omitted at this stage.}  The last step is to mix the wavefunctions $u_i$ with $u_j$, and $d_i$ with $d_j$, in a way that preserves orthogonality.  The updated wavefunctions have the form
\bea
u_i (x,s)&\ra& u'_i(x,s)= e^{i\th_1} \cos(\th_2) u_i(x,s) + \sin(\th_2) u_j(x,s) \non \\
u_j (x,s)&\ra& u'_i(x,s)= \cos(\th_2) u_j(x,s) -  e^{i\th_1} \sin(\th_2) u_i(x,s) \non \\
d_i (x,s)&\ra& d'_i(x,s)= e^{i\th_3} \cos(\th_4) d_i(x,s) + \sin(\th_4) d_j(x,s) \non \\
d_j (x,s)&\ra& d'_i(x,s)= \cos(\th_4) d_j(x,s) -  e^{i\th_3} \sin(\th_4) d_i(x,s) \ , \non \\
\eea
with the four angles $\th_{1-4}$ again determined numerically to minimize $\E$.  Note that all one-particle wave functions were normalized and orthogonal at initialization, and remain so after each relaxation step. This procedure continues for thousands of iterations, with the number required for convergence increasing with $N$ and the lattice size.

\subsection{Energy expectation values}
The 2D Hubbard Hamiltonian is, as usual,
\bea
H = -t \sum_{\langle x, y\rangle}\sum_s c_s^\dg(x) c_s(y) + 
U \sum_x c_\up^\dg(x)c_\up(x)c_\dn^\dg(x)c_\dn(x) \ . \non \\
\eea
Calculating $\E=\langle \Om|H|\Om\rangle$, for $|\Om\rangle$ in eq.\ \rf{Om}, is straightforward.  We need to define
\bea
\rho(x,s,s') &=& \sum_{i=1}^{M-N} \phi_i(x,s) \phi_i(x,s') \non \\
\rhot(x,s,s') &=& \sum_{i=1}^N b_i^2 (u^*_i(x,s) u_i(x,s') + d^*_i(x,s) d_i(x,s') ) \non \\
V_i(x) &=& u_i(x,\up) d_i(x,\dn) - u_i(x,\dn) d_i(x,\up) \non \\
W_i(x) &=& a_i b_i V_i(x)\ .
\eea
and also the operations
\bea
\Q[f(x,s)] &=&  f(x+\hat{e}_x, s) + f(x-\hat{e}_x, s) \non \\
& &+  f(x+\hat{e}_y, s) +  f(x-\hat{e}_y, s) \non \\
\R[f(x,s)] &=&  f(x+\hat{e}_x, s) + f(x-\hat{e}_x, s) \non \\
& & -  f(x+\hat{e}_y, s) -  f(x-\hat{e}_y, s) 
\eea
Where $f(x,s)$ may be an electron creation/destruction operator, or one of the one-particle wavefunctions 
$\phi_i(x,s), u_i(x,s), d_i(x,s)$.

Then the energy expectation value  is
\begin{widetext}
\bea
\E &=& -t \sum_{x,s} \sum_{i=1}^{M-N} \phi^*_i(x,s) \Q[\phi_i(x), s)]  
     -t \sum_{x,s} \sum_{i=1}^N b_i^2 \bigg\{ u^*_i(x,s) \Q[u_i(x,s)]  
 + d^*_i(x,s)  \Q[d_i(x, s)] \bigg\} \non\\ 
 & & + U\sum_x \bigg\{\rho(x,\up\up)\rho(x,\dn\dn) - \rho(x,\up\dn) \rho(x,\dn\up) ) 
 +\rhot(x,\up\up)\rhot(x,\dn\dn) 
          - \rhot(x,\up\dn)\rhot(x,\dn\up) \bigg\} \non \\
 & & + U\sum_x \bigg\{ \rho(x,\up\up)\rhot(x,\dn\dn) + \rho(x,\dn\dn) \rhot(x,\up\up) 
 -  \rho(x,\up\dn) \rhot(x,\dn\up) - \rho(x,\dn\up)\rhot(x,\up\dn) \bigg\}\non \\
 & & + U \sum_x \bigg\{ |\sum_i W_i(x)|^2 - \sum_i W_i^*(x) W_i(x) + 
         \sum_i (b_i^2-b_i^4)V_i^*(x) V_i(x)\bigg\} \ .
\label{energy}
\eea
\end{widetext}
This sum can be subdivided into a term which depends only on the $\phi_i(x,s)$, which includes part of the first line of 
\rf{energy}, as well as products of $\r \r$ (but not $\r \rhot$ and $\rhot \rhot$).  Since the $\phi$ wavefunctions are unchanged in the iterative procedure, the part of $\E$ which depends on them alone is subtracted, and we look for convergence, as described, until the remaining energy, involving the $N$ states in the neighborhood of the Fermi surface,  is stable up to the fourth digit.  The reason is that the total energy can be much larger than the part of the energy which depends on the $N$ states near the Fermi surface, and thus convergence of the difference to four decimal places is a much stronger condition than convergence of the total energy.

\subsection{Order parameters}

We define the momentum-space condensate
\bea
     P(k) &=& \langle \Om| c_\up(k) c_\dn(-k) |\Om \rangle \non \\
             &=& \sum_{i=1}^N a_i b_i (u_i(k,\up) d_i(-k,\dn) - u_i(-k,\dn)d_i(k,\up)) \ ,
\label{Pk}
\eea
and the d$_{x^2-y^2}$ parameter \cite{Scalettar2}
\bea
\D_1 &=& {1\over L^2} \sum_x \D_1(x) \non \\
&=& {1\over L^2} \sum_x \langle \Om| c_\up(x)\R[c_\dn(x)] |\Om \rangle \non \\
&=& {1\over L^2} \sum_x \sum_{i=1}^N a_i b_i \bigg\{ \R[u_i(x,\up)] d_i(x,\dn) \non \\
& & \qquad  \qquad - \R[d_i(x,\up)] u_i(x,\dn) \bigg\} \non \\
&=& 2 \sum_k P(k) (\cos(k_x) - \cos(k_y)) \ ,
\label{dxy}
\eea
which of course has maximum amplitude in the ideal case that $P(k) \propto \cos(k_x) - \cos(k_y)$.  Likewise,
the s-wave order parameter is
\bea
\D_2 &=& {1\over L^2} \sum_k \sum_{i=1}^N  a_i b_i(u_i(x,\up) d_i(x,\dn) - u_i(x,\dn)d_i(x,\up)) \non \\
&=& 2 \sum_k P(k) \ ,
\eea
and we denote a variant order parameter s$_{x^2+y^2}$
\bea
\D_3 &=& {1\over L^2} \sum_x \langle \Om| c_\up(x) \Q[c_\dn(x)] |\Om \rangle \non \\
&=& {1\over L^2} \sum_x \sum_{i=1}^N a_i b_i \bigg\{ \Q[u_i(x,\up)]d_i(x,\dn) 
 + \Q[d_i(x,\up)] u_i(x,\dn) \bigg\} \non \\
&=& 2 \sum_k  P(k) (\cos(k_x) + \cos(k_y)) \ .
\eea

Assuming d-wave condensation occurs, we would like to know if the magnitude found for $\D_1$ is actually
non-negligible.  Suppose the wavefunctions $u_i,d_i$ are spatially extended rather than localized.  Then products of
$u_i d_i$ are $O(1/L^2)$.  Next we have to guess number of levels denoted by $i$ for which the product
$a_i b_i$ is non-negligible.  Call this fraction $q$, and then, of the levels which contribute, denote the average magnitude of $a_i b_i$ as $\overline{ab}$.  So the back-of-the-envelope estimate is that $\D_1$ is of order $q \overline{ab}$.  If we just guess that perhaps one percent of O($L^2$) levels contribute, with $\overline{ab} \sim 0.1$, then a 
magnitude of $\D_1 \sim O(10^{-3})$ seems reasonable, although we should not be surprised if the actual value would be different by an order of magnitude.  We will report the values of $|\D_{1,2,3}|$ in the
range $2\le U/t \le 8.5$ and density $f$ up to half filling, in the range $0.4 \le f \le 1$.

\section{Results}

   In principle both $P(k)$ and the order parameters $\D_{1,2,3}$ are complex.  Below we show either the
modulus or real value in parameter space.  We have studied the imaginary parts also, but these are in general a factor of at least five smaller than the real part.

   In the simulations we set $t=1$ ($U/t=U$) and, at each $U,f$, relax to thirty different local minima starting from thirty different initial states, and compute average values  and error bars from this set.  Next-to-nearest couplings are neglected ($t'=0$).  
   
   One might ask why we do not simply select the state with the lowest energy.  Our view, put forward in \cite{Matsuyama:2022kam}, is that since the Hartree-Fock approximation to the 2D Hubbard model has very many consistent solutions, all quite close in energy density, this might also be a feature of the exact Hubbard model, meaning that are a large number of states with very nearly the same energy as the true ground state, reminiscent of a spin glass.  In a spin glass it is misleading to attribute a special significance to one unique state of lowest energy,
since there are a vast number of states which differ from the lowest energy state by insignificant differences
in energy and, presumably, in physical properties. That is certainly the case for the local energy minima generated by our procedure.  For this reason we do not concentrate on a local minimum with the lowest energy out of thirty runs; more runs will always generate, eventually, some state with still lower energy.  We believe it is preferable, in reporting order parameters, to average over thirty very similar local minima, which would minimize the importance of any outliers.

\begin{figure}[htb]
 \center
 \includegraphics[scale=0.7]{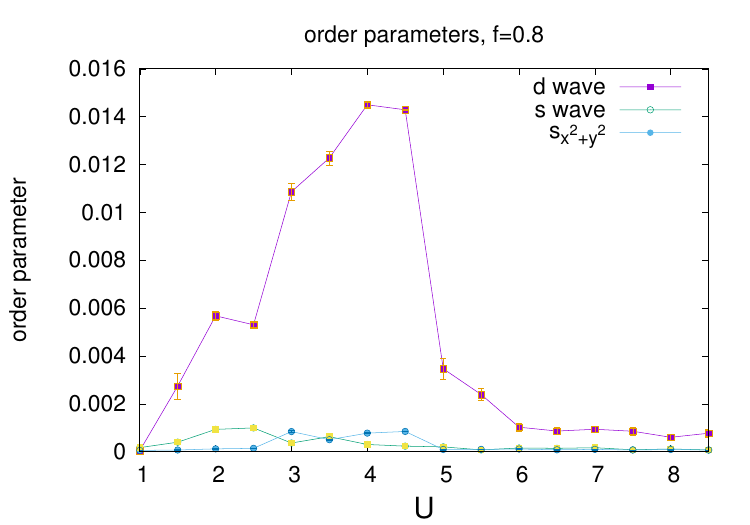}
 \caption{Condensates of s and d-wave type vs.\ $U$ in the local minimum state at density $f=0.8$, $10\times 10$ lattice volume. The real part is shown.} 
 \label{dwave}
 \end{figure}
   
   We begin with results, in Fig.\ \ref{dwave}, for the real part of condensates $\D_{1,2,3}$ vs.\ $U=1-8.5$ at a fixed density $f=0.8$ with $N=20$, on a $10\times 10$ lattice.  Obviously the s-wave ($\D_2$) and s$_{x^2+y^2}$ ($\D_3$) condensates are negligible compared to the d-wave condensate ($\D_1$) in the range of moderate interaction $2\le U \le 5$.  The bars shown on d-wave data points represent the standard deviation, not the standard error, and are an indication of the deviation in the order parameter among local minima (obtained with independent starting points) at the same coupling and density.  
   
\begin{figure}[h!]
\subfigure[~]  
{   
 \label{d3d}
 \includegraphics[scale=0.32]{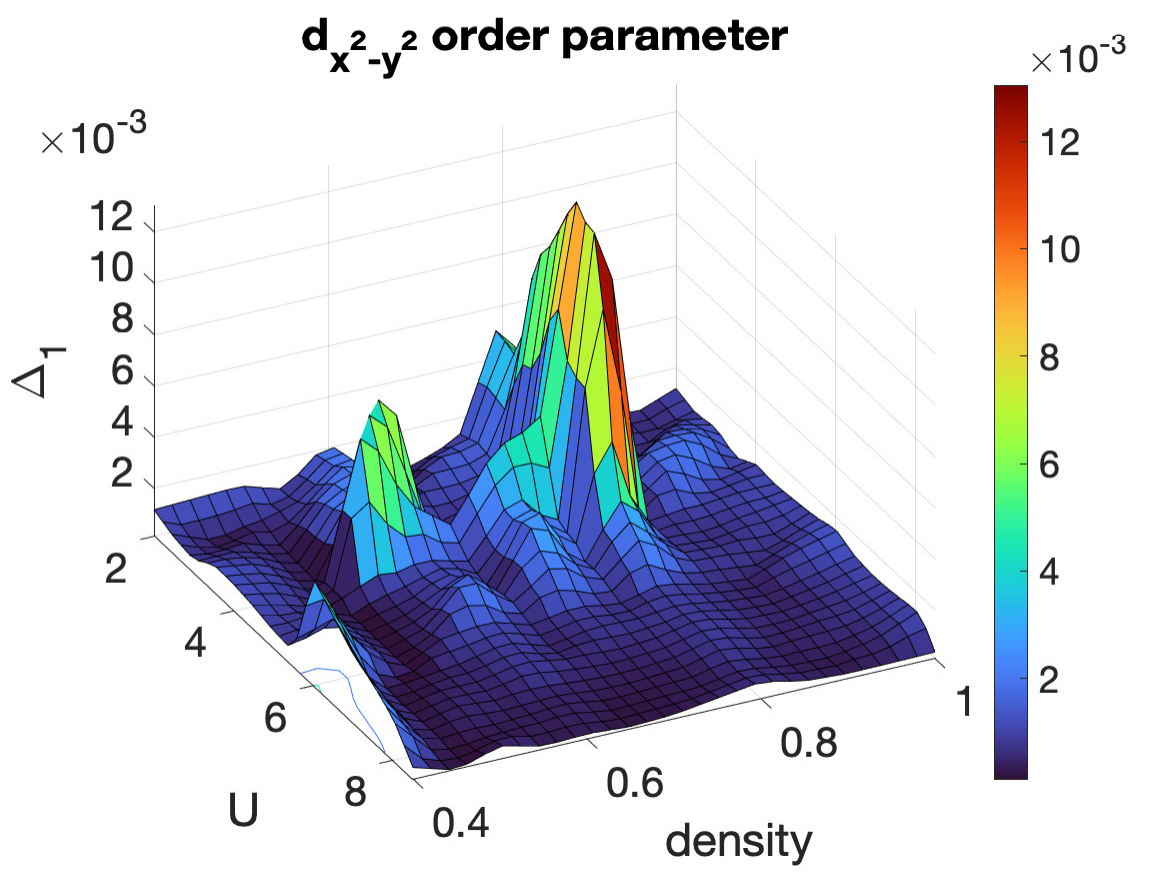}
}
\subfigure[~]  
{   
 \label{topview}
 \includegraphics[scale=0.32]{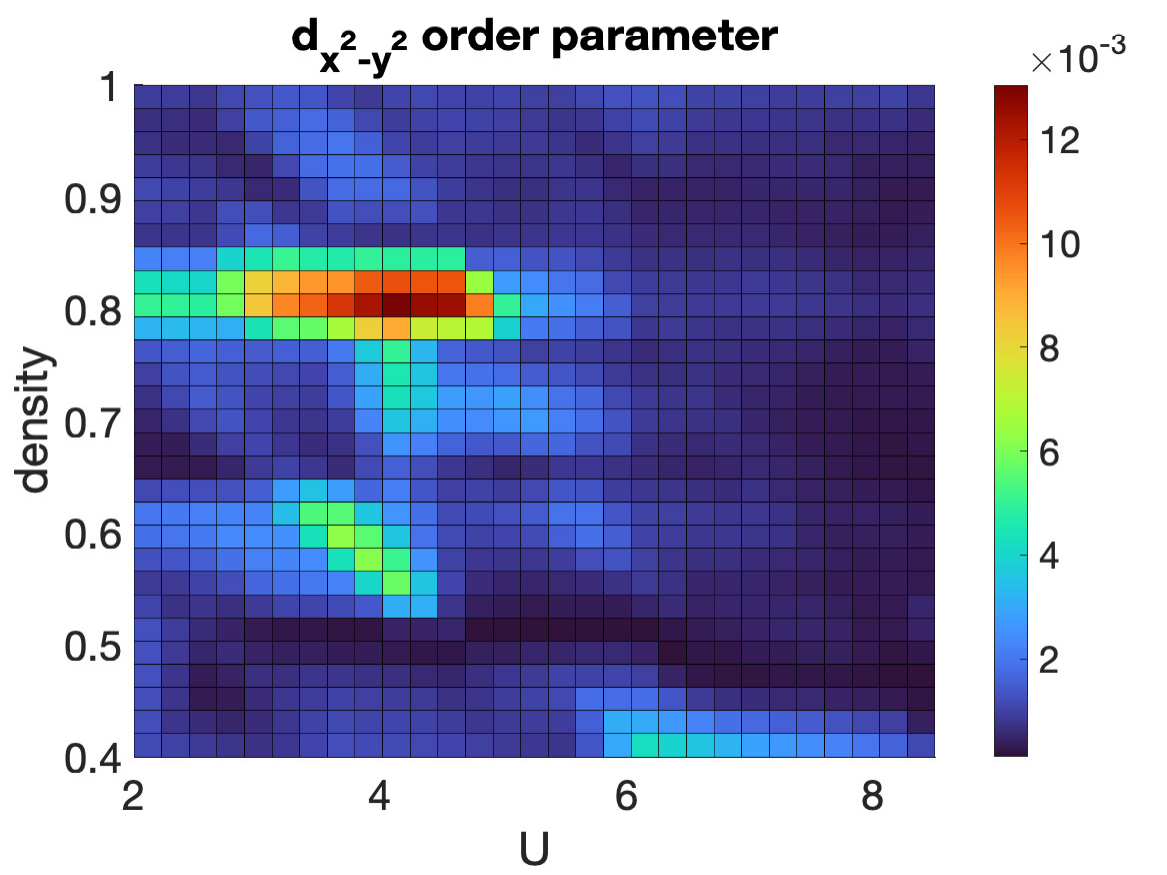}
}
\subfigure[~]
{   
 \label{s3d}
 \includegraphics[scale=0.32]{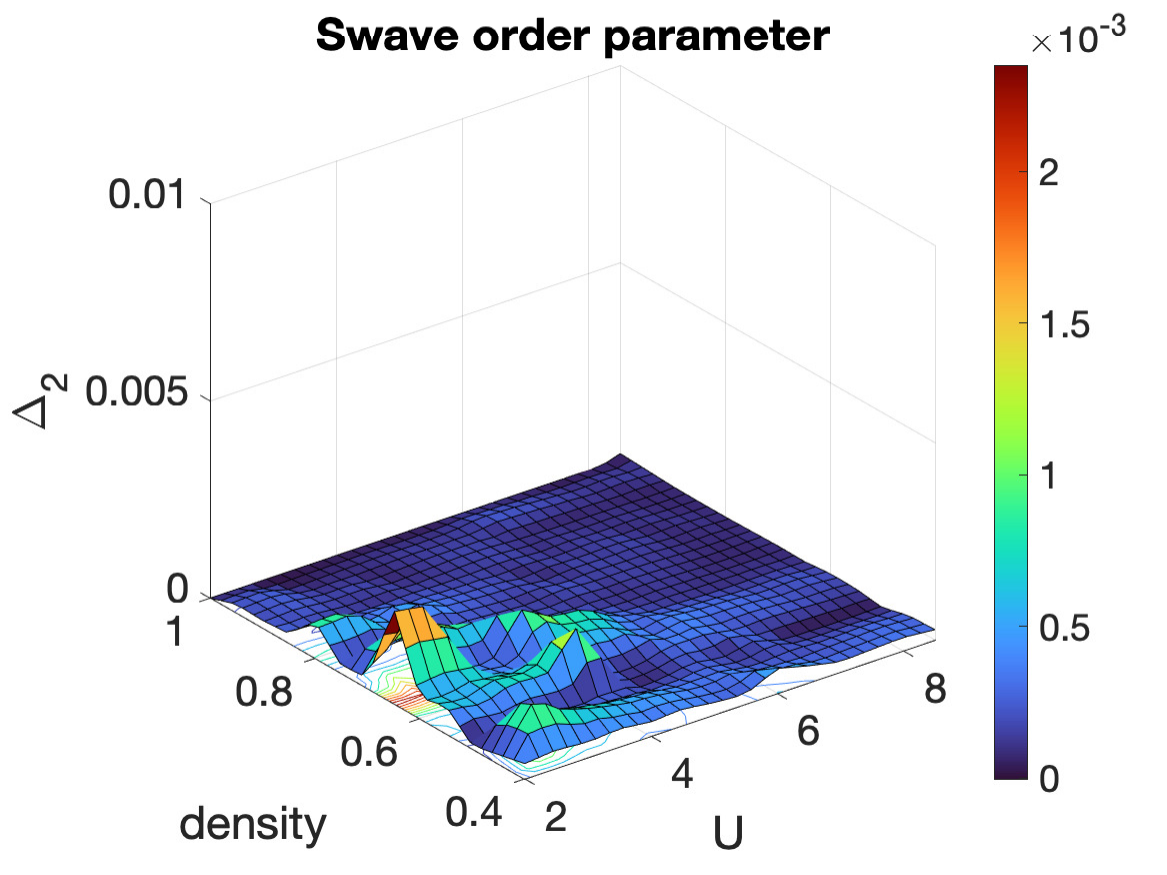}
}
\subfigure[~]
{   
 \label{v3d}
 \includegraphics[scale=0.32]{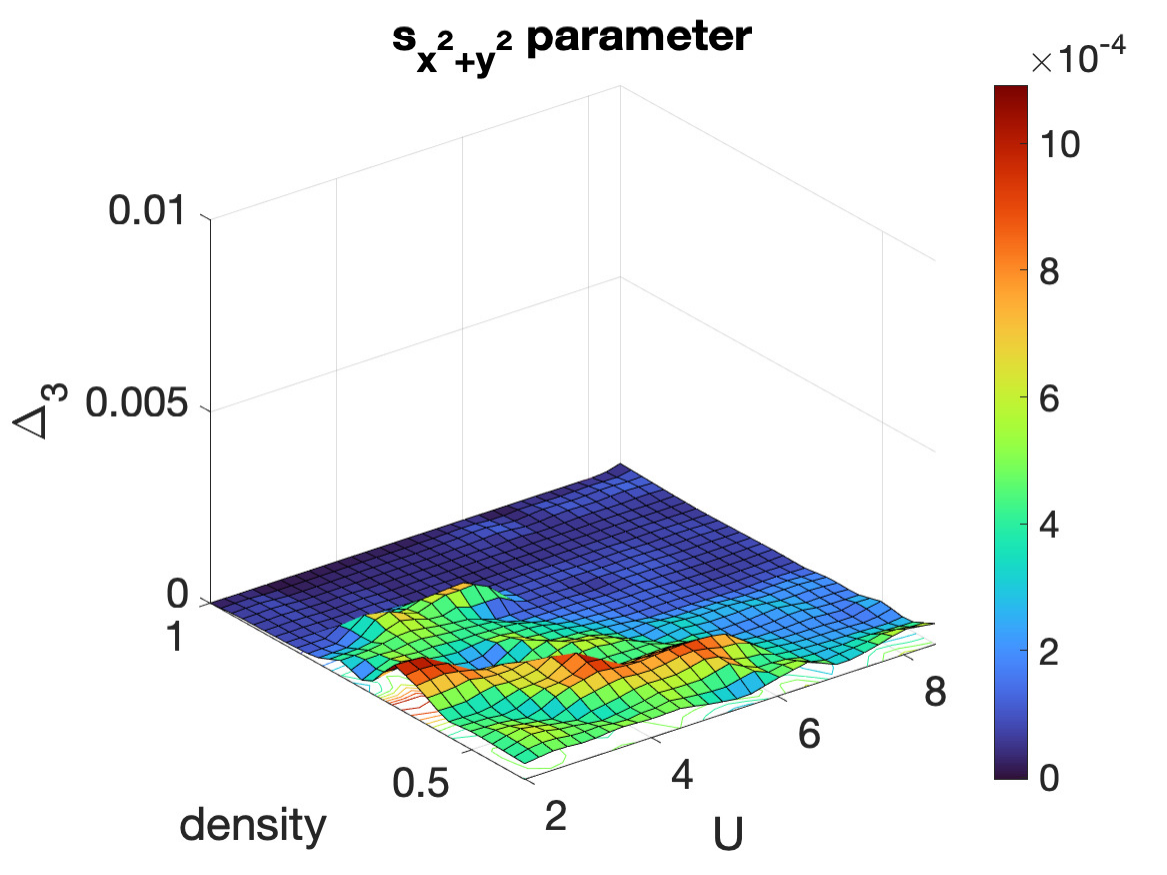}
}
\caption{3d view of the real part of (a) the d-wave order parameter; (c) the s-wave order parameter; and
(d) the $s_{x^2+y^2}$ order parameter defined in the text, in the range of couplings $U$ and 
densities $f$ shown.  Subfigure (b) is a ``top down'' display of the d-wave condensate with the height
of the condensate indicated by color, rather than z-axis position, to help delineate the region of the d-wave peak.} 
\label{view3d}
\end{figure}

    The d-wave condensate for densities in the range ${0.4\le f \le 1}$ is shown in
Figs.\ \ref{d3d} and \ref{topview}, again at $N=20$ on a $10\times 10$ lattice.  This should be compared 
with the results for $\D_2$ and $\D_3$ shown, on the same scale and 
parameter range, in Figs.\ \ref{s3d} and \ref{v3d} respectively.  From the top down view in Fig.\ \ref{topview} we see that the d-wave order parameter peaks in the region of coupling $2\le U \le 5$ and density $0.78\le f \le 0.85$, while the s-wave and s$_{x^2+y^2}$ are negligible by comparison, and show no significant peaks in the condensate.  It should be noted that the 3D figures were created by graphic software which interpolates values from the coarse discretization of the $U-f$ plane to produce a surface which, for the sake of clarity, is smoother in appearance than what would be obtained from, e.g., a simple histogram of the data.
The actual discretization in most cases is $\D U = 0.5, \D f = 0.05$.

   The maximum of the condensation peak is in the neighborhood of $U=4, f=0.8$.  There is another recent work, using quite different methods, that also finds a condensate near these parameters \cite{Pfaffian2}.
  
\begin{figure}[t!]
  \subfigure[~]{
 \includegraphics[scale=0.25]{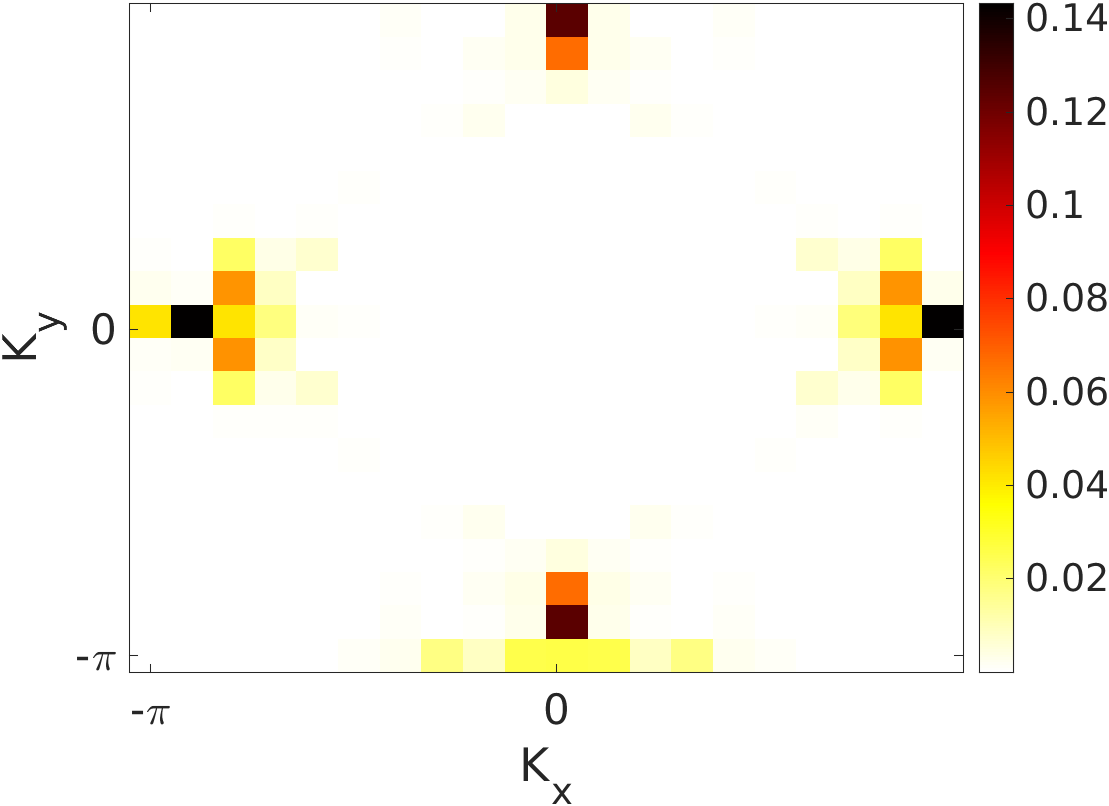}
 \label{absorder4}
 }
  \subfigure[~]{
 \includegraphics[scale=0.25]{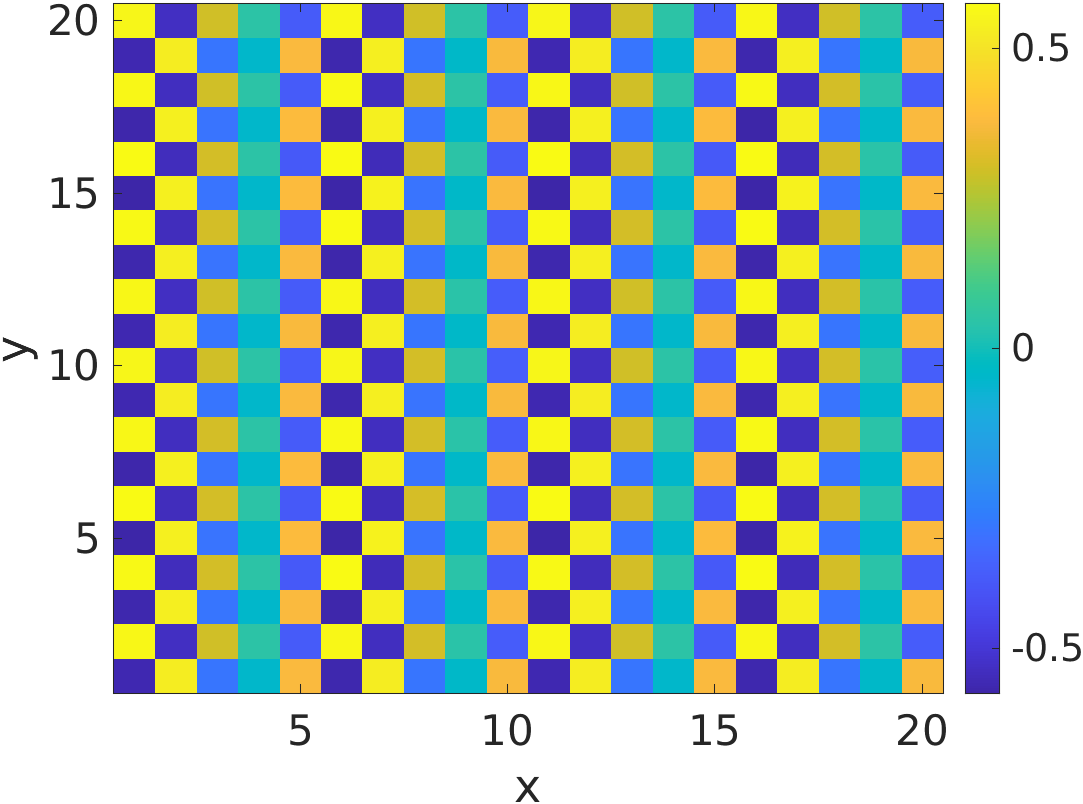}
 \label{geospin4}
 }
  \subfigure[~]{
  \includegraphics[scale=0.25]{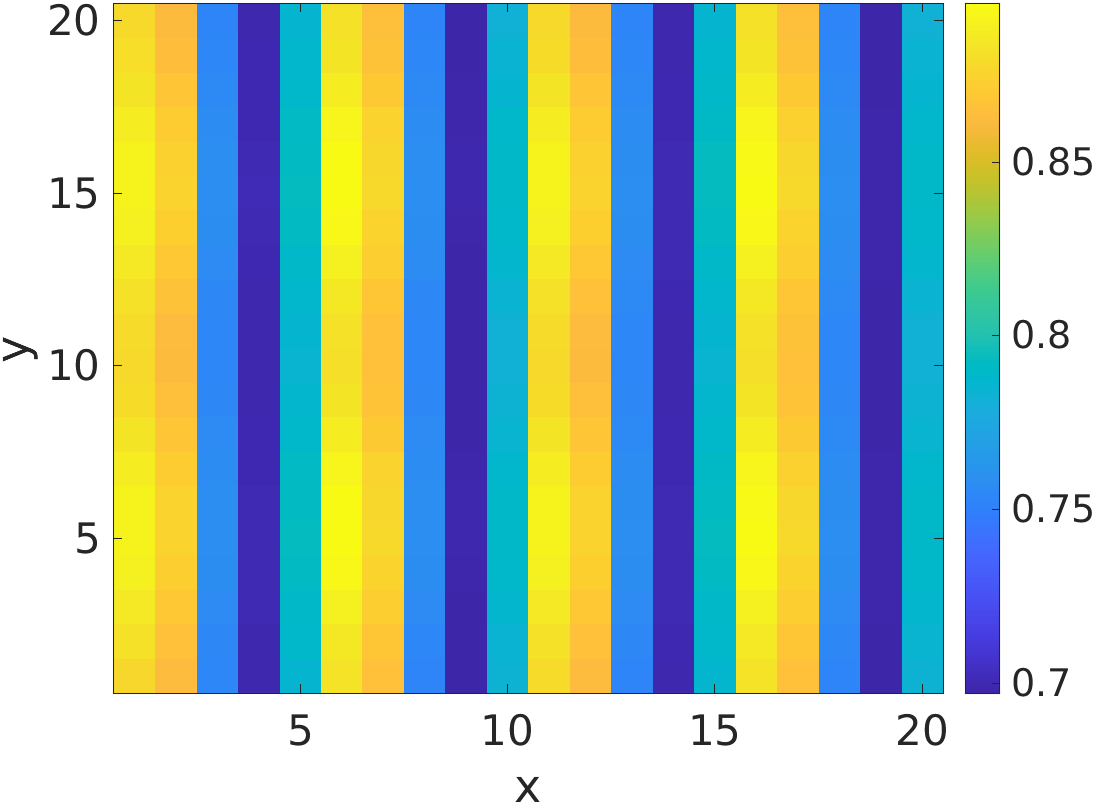}
 \label{cdw4}
 }
 \subfigure[~]{
  \includegraphics[scale=0.25]{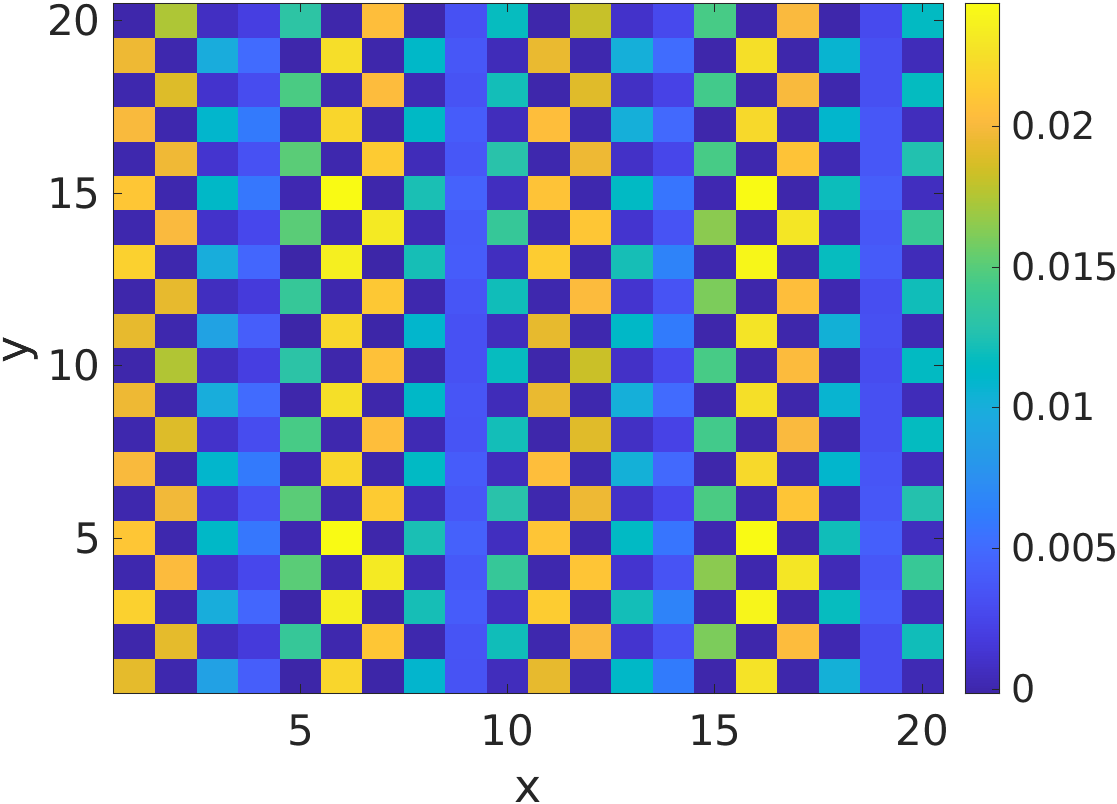}
 \label{order4}
 }
 \caption{ (a) Modulus of the momentum space condensate $P(k)$. (b) Spin density $D(x)$. (c) Charge (number density) $C(x)$. (d) The real part of the condensate spatial distribution $\D_1(x)$, which in this case is positive at all sites.  All figures from single configurations at $U = 4, f = 0.8$ on a $20\times 20$ lattice with ${N=80}$.  Other configurations display horizontal, rather than vertical stripes. } 
 \label{geometry}
 \end{figure}

     Define the spin density
\beq
      D(x) = \rho(x,1,1) + \rhot(x,1,1) - \rho(x,2,2) - \rhot(x,2,2) \ ,
\eeq
charge (or electron number) density
\beq
      C(x) = \rho(x,1,1) + \rhot(x,1,1) + \rho(x,2,2) + \rhot(x,2,2) \ ,
\eeq
and the modulus $|P(k)|$ and $\D_1(x)$ were defined in \rf{Pk} and \rf{dxy}. These quantities are illustrated in Fig.\ \ref{geometry} in the region of the d-wave condensate peak in $\D_1$ on a $20\times 20$ lattice volume at $U=4, f=0.8, N=80$.  Fig.\ \ref{absorder4} shows a feature which is very typical in the condensate region, namely that $P(k)$ is largest in magnitude 
when $\cos(k_x)-\cos(k_y)$ is largest in magnitude, and negligible in the region where $\cos(k_x)-\cos(k_y)$ vanishes (the nodes).  Fig.\ \ref{geospin4} shows a stripe pattern typical of the spin density $D(x)$ in the stripe phase, while \ref{cdw4} is a plot of the charge density.  Finally, Fig.\ \ref{order4}  displays the condensate density in position space.  We note that all of these observables show similar geometric patterns, which seems to be an instance of intertwined order \cite{Fradkin}.
  
\clearpage 
 
 \begin{figure}[htb]
 \subfigure[~]{
 \includegraphics[scale=0.3]{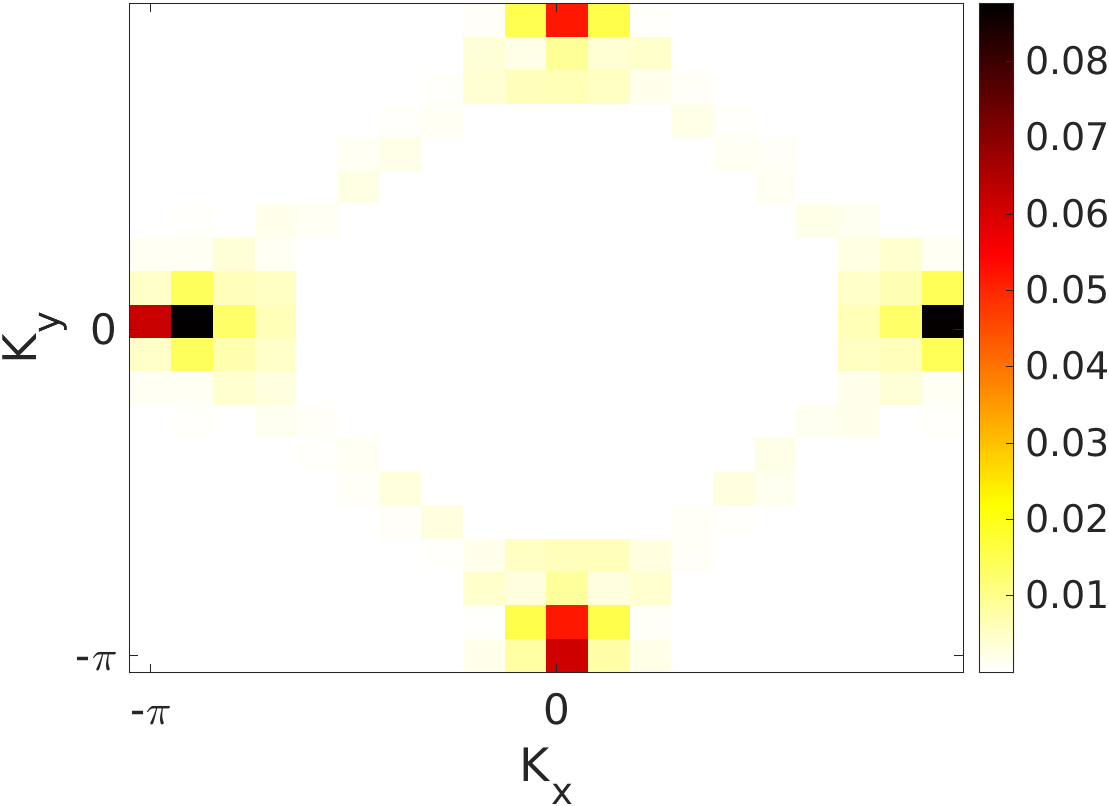}
 \label{absorder085}
 }
 \subfigure[~]{
  \includegraphics[scale=0.3]{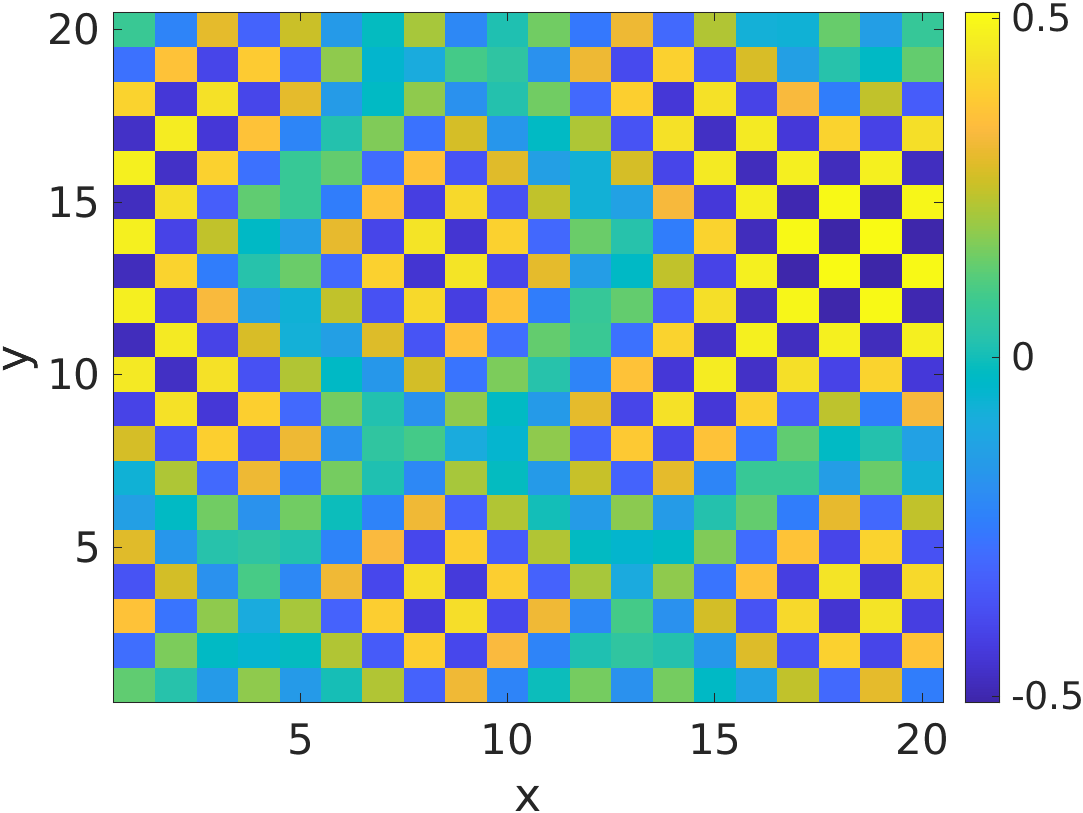}
 \label{geo085}
 }
  \subfigure[~]{
  \includegraphics[scale=0.3]{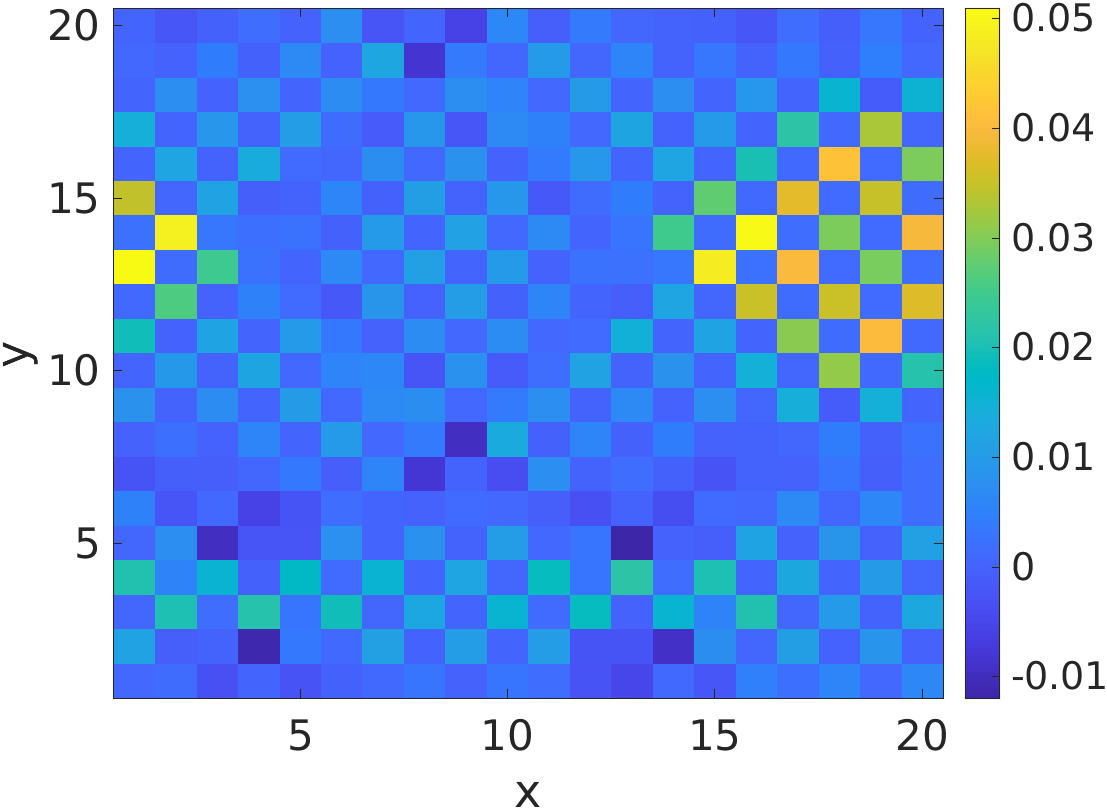}
 \label{order085}
 }
 \caption{ Similar to Fig.\ \ref{geometry}, but at $U=3, f=0.85$.
 (a) Modulus of the momentum space condensate $P(k)$. 
 (b) Spin density $D(x)$. (c) The real part of the condensate spatial distribution $\D_1(x)$.  Stripes are less
 evident, and the condensate is not positive everywhere.  All figures again taken from a single configuration on a $20\times 20$ lattice with $N=80$. } 
  \label{f085}
 \end{figure}

Of course the fact that $\D_1$ is significantly different from zero in a certain region of $U-f$ parameter
space is not a guarantee that  the condensate $P(k)$ closely adheres to the $\cos(k_x)-\cos(k_y)$ form, as we already see in plotting $|P(k)|$ in the condensate region,  cf.\ Fig.\ \ref{absorder4} at $U=4$  and Fig.\ \ref{absorder085} at $U=3$.  But while details can vary widely from one configuration to the next, what we do see clearly is the suppression of $|P(k)|$ at the nodes $k_x= \pm \pi/2, k_y =\pm \pi/2$ and maximization at the peaks $k=(0,\pm \pi), (\pm \pi,0)$ of $|\cos(k_x)-\cos(k_y)|$ in the region of the condensate peak.

\begin{figure}[htb]
\includegraphics[scale=0.3]{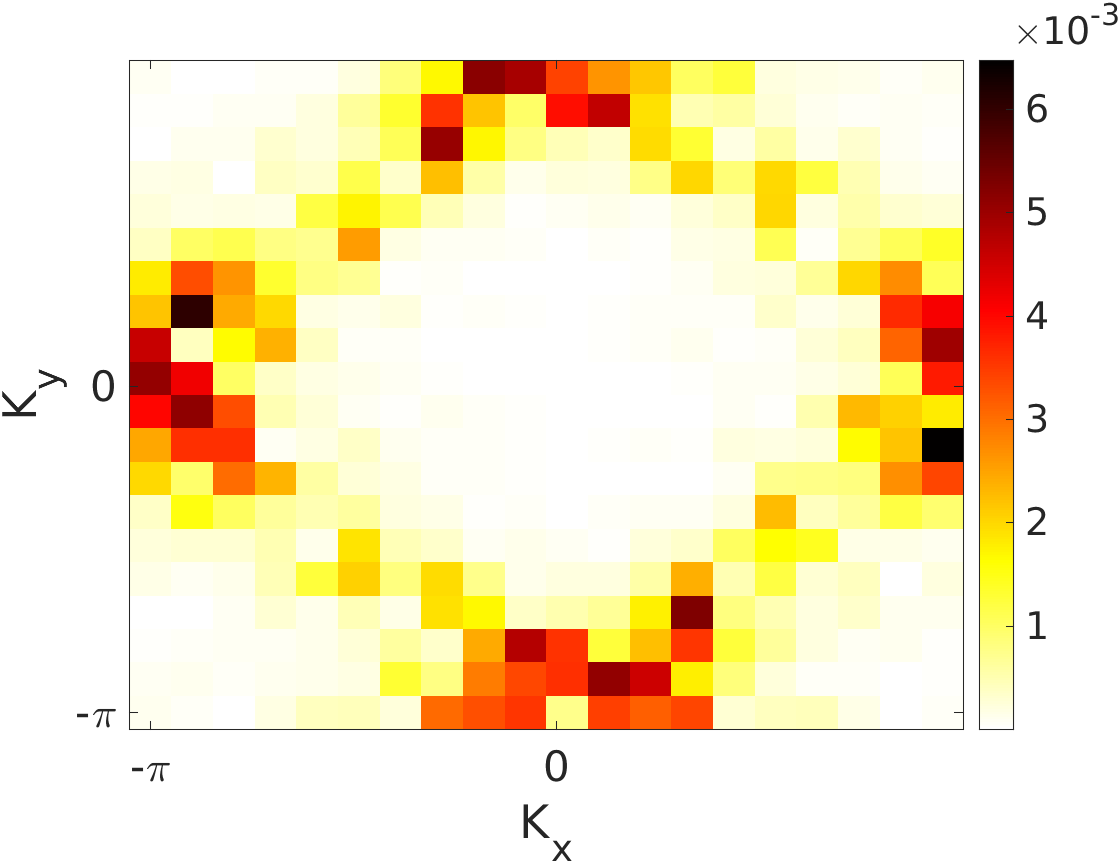}
\caption{$|P(k)|$ at a moderately strong coupling of $U=8,f=0.8$ on a $20\times 20$ lattice.  A d-wave pattern
is evident, although the amplitude is greatly reduced compared to the peak region.}
\label{absorder24}
\end{figure}

We see from Fig.\ \ref{view3d} that the d-wave order parameter $\D_1$, while still at least an order of magnitude greater  than $\D_2$ and $\D_3$, is greatly reduced away from the peak values.
As an example, in Fig.\ \ref{absorder24}, we display $|P(k)|$ at $U=8$ and $f=0.8$ on a $20\times 20$ lattice, which is well away from the peak, and can be compared with the corresponding plot in Fig.\ \ref{absorder4}.  Perhaps it is
reasonable to dismiss a finite $\D_1$ away from the peak values as ignorable.  But taking the non-zero value of $\D_1$ at face value, it is also possible that we have a weak d-wave condensate quite far from the peak,
and into the strong-coupling regime.

 \begin{figure}[htb]
 \center
 \includegraphics[scale=0.5]{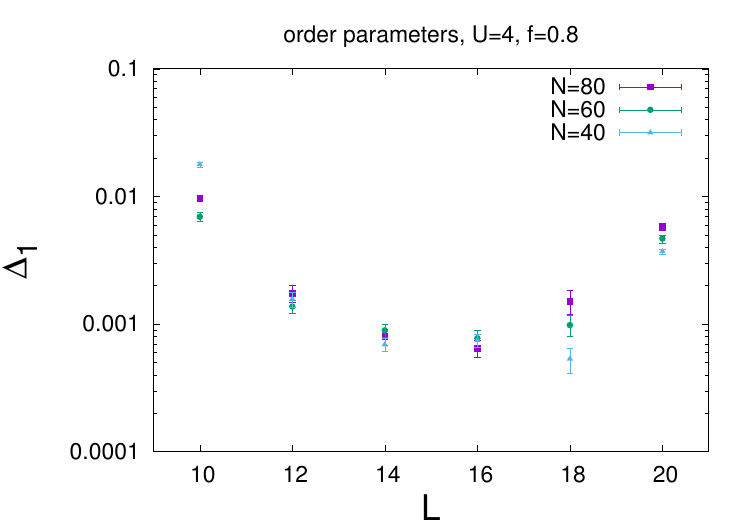}
 \caption{Condensate $\D_1$ vs. lattice extension $L$ at $U=4$, density 0.8, and $N=40,60,80$.} 
 \label{NvsL}
 \end{figure}
 
 \begin{figure}[htb]
 \subfigure[~]{
 \includegraphics[scale=0.25]{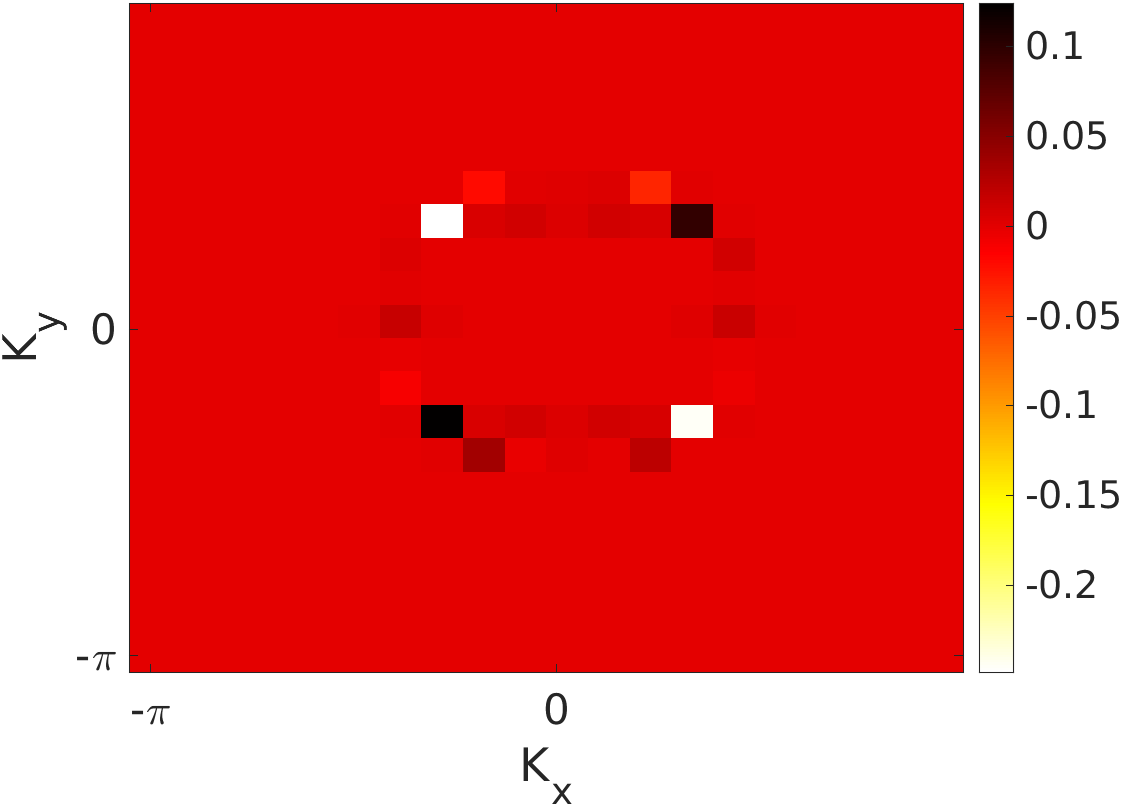}
 \label{dxyU4}
 }
 \subfigure[~]{
  \includegraphics[scale=0.25]{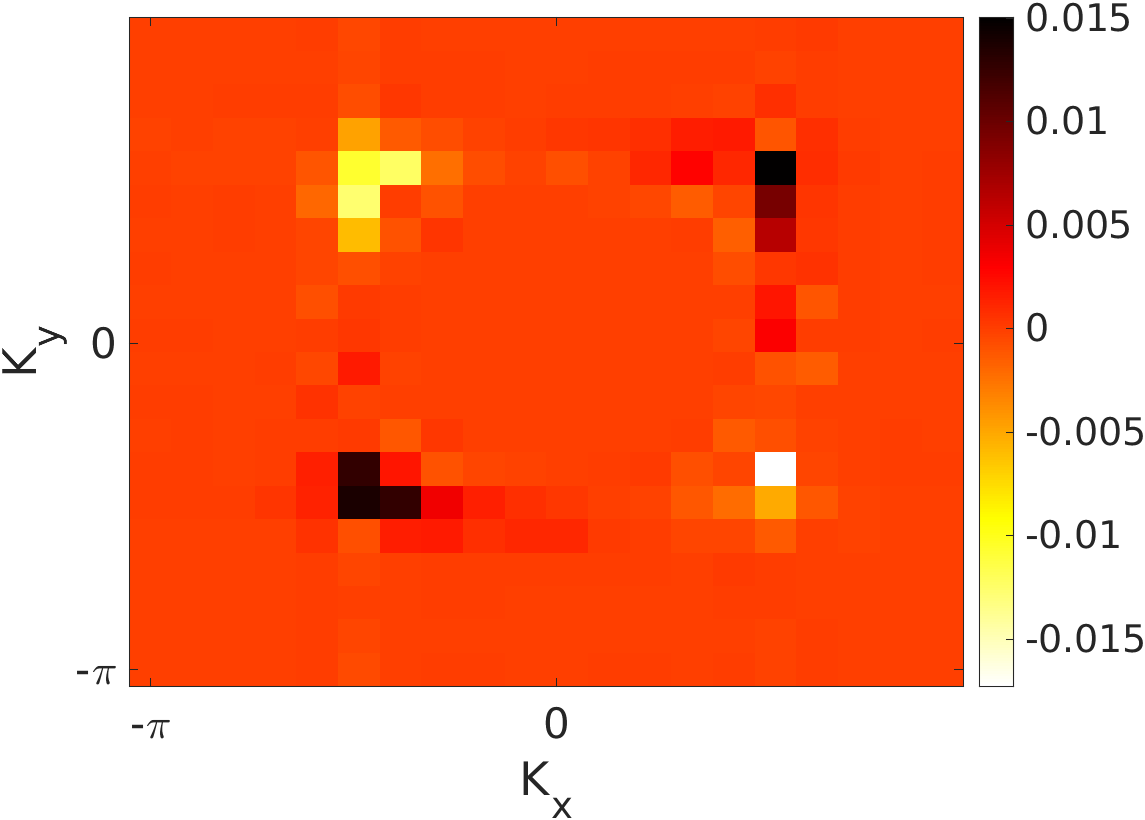}
 \label{dxyU8}
 }
 \caption{Evidence of d$_{xy}$ condensation, in the overdoped regime, from plots of the real part of $P(k)$. (a) $U=4, f=0.3$.  (b) $U=8, f=0.5$.   The plot subfigures (a) and (b) are derived from a single BCS state on a $20\times 20$ lattice in each case.}
 \label{dxyfigs}
 \end{figure}
 
   In order to perform calculations in a reasonable time at many values of coupling $U$ and density $f$, our computations in the $f-U$  plane were carried out on relatively small $10\times 10$ lattices.  A natural question is how the strength of the d-wave condensate $\D_1$ varies with lattice size.  To investigate this question we computed the  condensate at $U=4$ and density $0.8$, which is where the condensate peaks on the $10\times 10$ lattice, on $L\times L$ lattices with $L = 10-20$, and $N=40,60,80$ on each lattice.  The results are shown in Fig.\ \ref{NvsL}.  The amplitude of the condensate in this range of $N$ is only weakly dependent on $N$, but it is curious that  there are substantial finite size effects in the magnitude.

\subsection{$d_{xy}$ condensation in the overdoped regime}

We have so far concentrated on the underdoped, high density region, where the d-wave condensation of the expected $d_{x^2-y^2}$ symmetry is
dominant and peaked in the region shown in Fig.\ \ref{view3d}.  However, further investigation in the overdoped, low electron density region indicates condensation of the $d_{xy}$ type. Evidence of  $d_{xy}$ condensation in the overdoped regime is displayed at $U=4, f=0.3$ and $U=8, f=0.5$ on $20\times 20$ lattices
in Figs.\ \ref{dxyU4} and \ref{dxyU8} respectively.  In these figures we plot the real part of $P(k)$,  and it is seen that peaks in the magnitude of $P(k)$, and their relative signs, agree with what is expected in $d_{xy}$ symmetry.  It is of interest that the existence of $d_{xy}$ condensation in the overdoped Hubbard model at weak to intermediate couplings, specifically including $U=4,f=0.3$, has been seen previously by Deng et al.\  \cite{Deng}using a combined diagrammatic and Monte Carlo approach.  We are not aware of other results in the literature at stronger couplings in the overdoped region.

 \begin{figure}[htb]
 \includegraphics[scale=0.35]{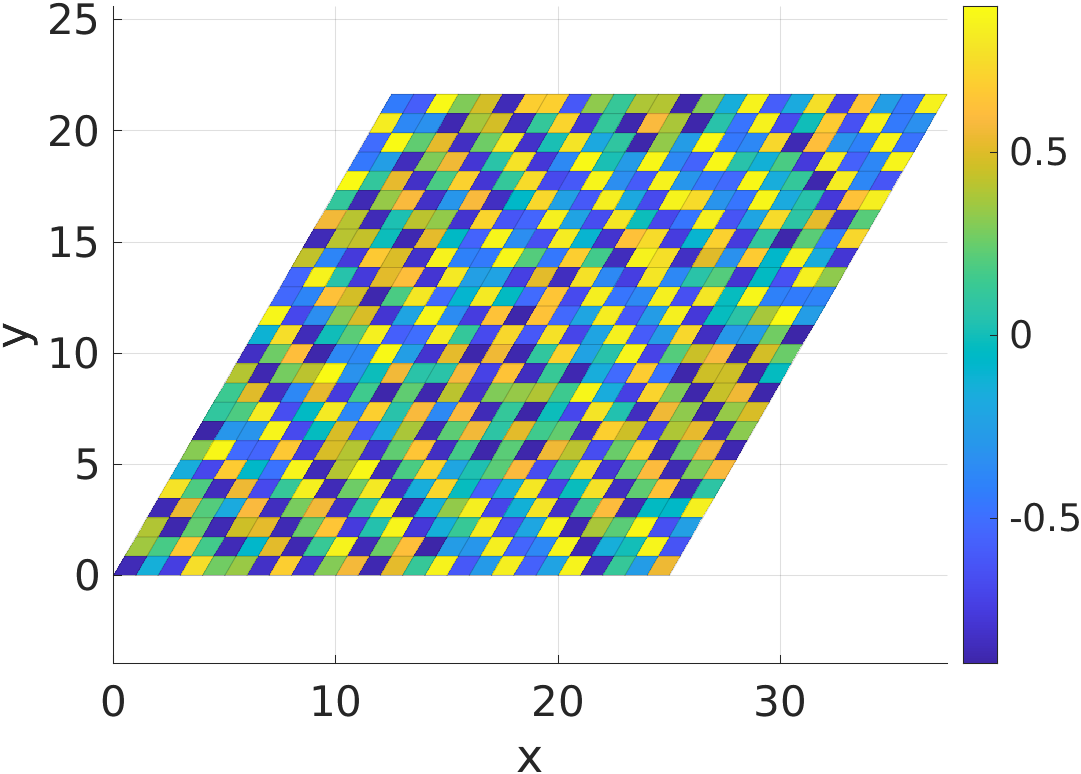}
 \caption{A plot of the spin density $D(x)$ on the triangular lattice at $U=8$ and half-filling.  The well-known checkerboard pattern which is seen on a square lattice at half-filling is absent in the case, due to lattice frustration of antiferromagnetic order.}
 \label{geotri}
 \end{figure}
\begin{figure*}[t]
 \subfigure[~$P_{max}$]{.
 \includegraphics[scale=0.35]{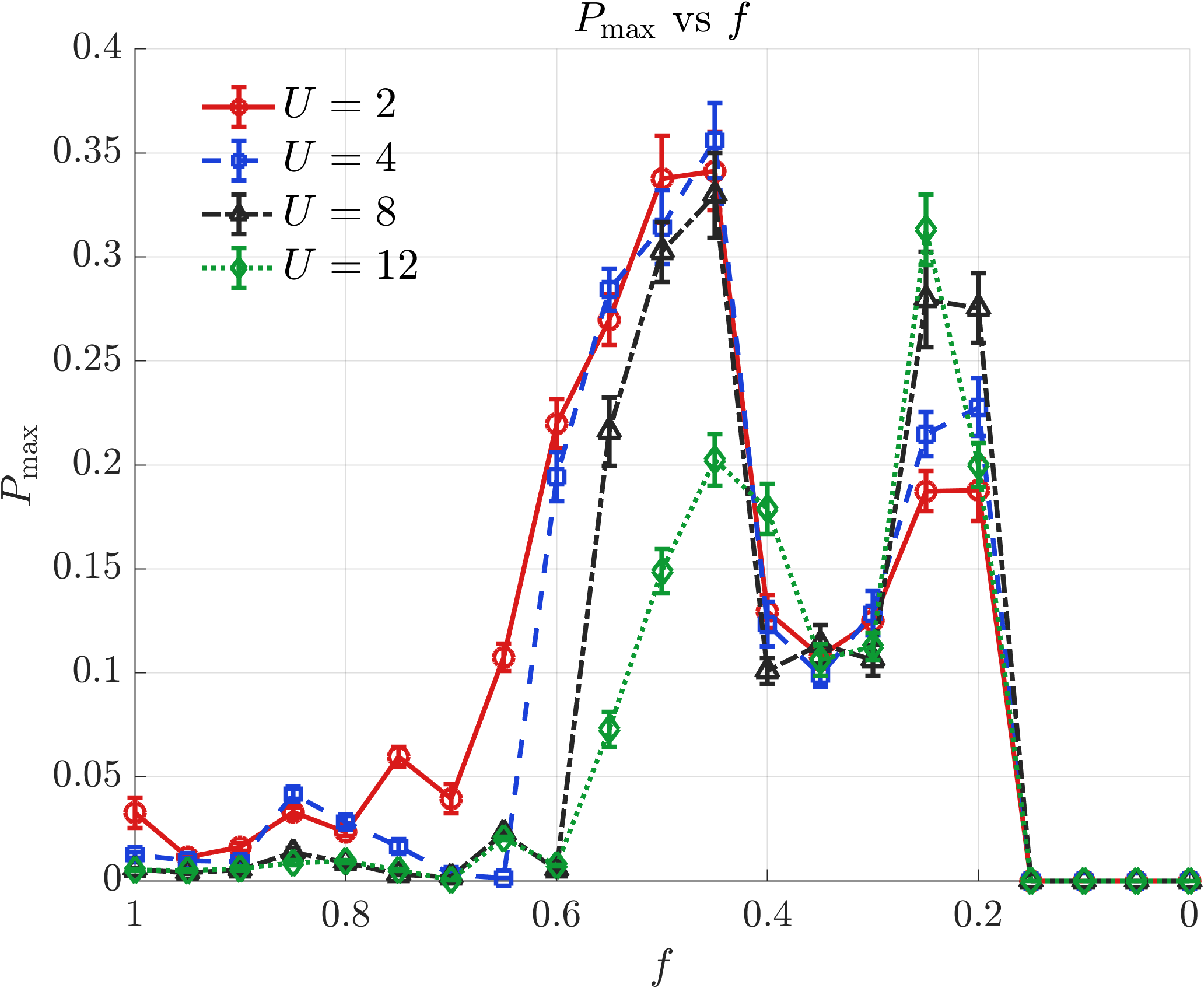}
 \label{Pmax}
 }
 \hspace{50pt}
 \subfigure[~$P_{sum}$]{
   \includegraphics[scale=0.35]{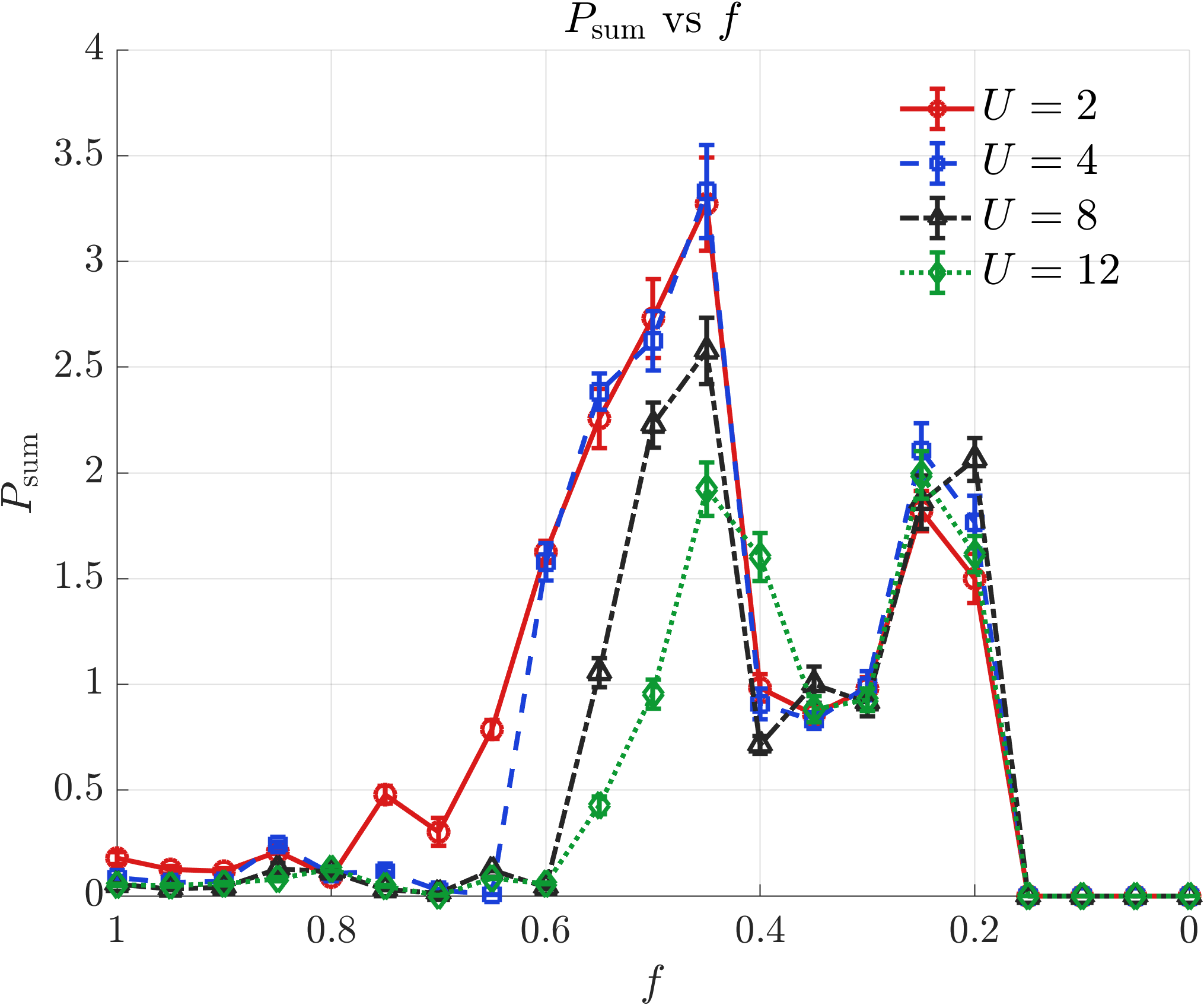}
 \label{Psum}
 }
 \caption{A plot of (a) the maximum value of $P(k)$, and (b) the sum of $P(k)$ over $k$, discarding values
 less than $10^{-3}$.  Data is taken at the electron densities $f$ and couplings $U$ shown, on a $12\time 12$ lattice.}
  \label{Ptriangle}
 \end{figure*}

\subsection{Condensation on a triangular lattice}

    The Hubbard model on a triangular lattice has not received quite as much attention as square lattices in the literature (for a sample on this topic, however, see \cite{Arovas_2022,Qin_review,Qin_2022}) and references therein).  One may think of a square lattice as an array of plaquettes with lattice sites at the corners. By simply adding one hopping term across one of the diagonals of each plaquette to the Hubbard Hamiltonian,
equal in strength ($t=1$) to the nearest-neighbor hopping terms, the system becomes a triangular lattice, although on a square lattice the three sides of each triangle are not equal in length in real space.  Although triangle side lengths are irrelevant to the Hamiltonian as just described, all results obtained with sites on a square lattice can easily be mapped to the appropriate lattice corresponding to an array of equilateral triangles, with the sites of the Fourier transformed data mapped to the corresponding reciprocal lattice.
Explicitly, if the coordinates on a square array and in the corresponding Fourier transform space are denoted 
by a pair of integers $(i,j)$, then these coordinates are mapped
to a system of equilateral triangles in real space with sites in Cartesian $x,y$ coordinates 
\beq
x^{ij} = i + \oh j   ~~~,~~~ y^{ij} = {\sqrt{3}\over 2} j~~~~~~ (1 \le i,j \le L)
\eeq
 Likewise, if the discrete Fourier transform of data obtained on a square lattice with a diagonal hopping term are located at momenta specified by integers 
$(m,n)$, then these sites are mapped to to the reciprocal lattice by
\bea
    & & k^{mn}_x = \pi {m\over L} ~~~,~~~k^{mn}_y = {\pi \over \sqrt{3}} {2n - m\over L}  \non \\
    & & (-L \le m,n \le L-1)
\eea
In both cases, a periodic square lattice is mapped onto (different) periodic parallelograms in real and reciprocal lattices, respectively.  On a square lattice with only nearest-neighbor couplings, antiferromagnetic order at
half-filling shows up as a checkerboard pattern in the spin distribution (see, e.g.\, Fig.\ 10(d) in \cite{Matsuyama:2022kam}).  On a triangular lattice, which frustrates this arrangement, no checkerboard is seen at half-filling.  Instead we see arrangements such at the one shown in Fig.\ \ref{geotri} at $U=8, f=1$.

\begin{figure*}[htb]
 \subfigure[~real, $f=0.25$]{.
 \includegraphics[scale=0.2]{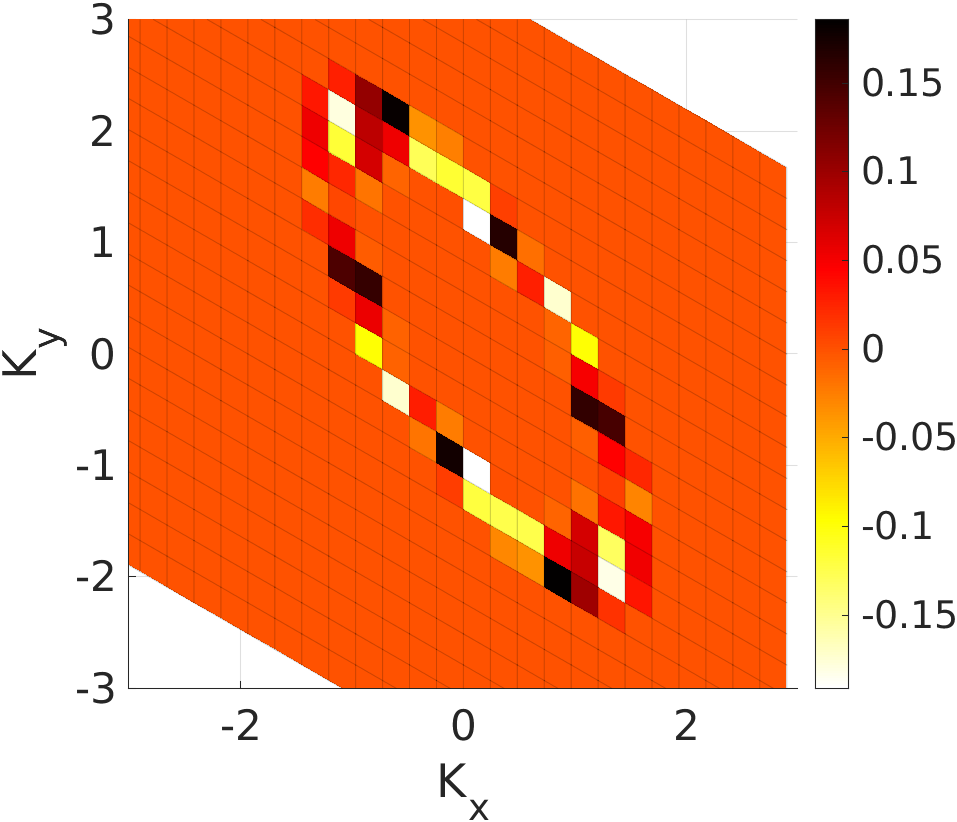}
 \label{real4f25}
 }
 \subfigure[~imag, $f=0.25$]{
   \includegraphics[scale=0.2]{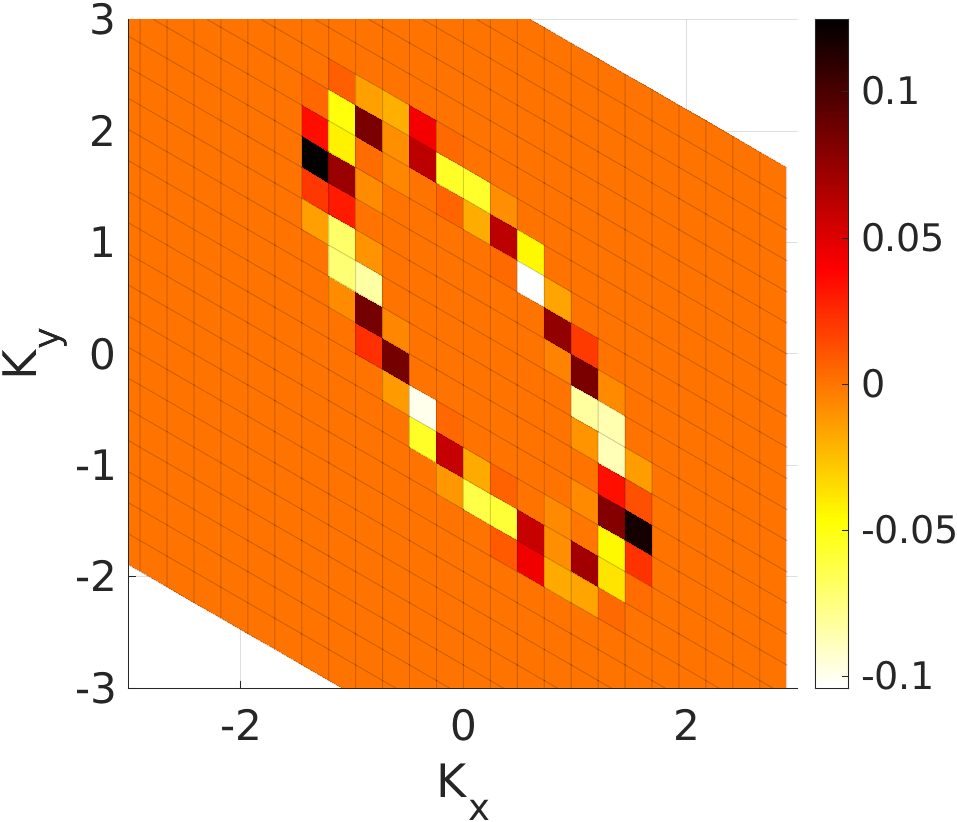}
 \label{imag4f25}
 }
  \subfigure[~real, $f=0.45$]{.
 \includegraphics[scale=0.2]{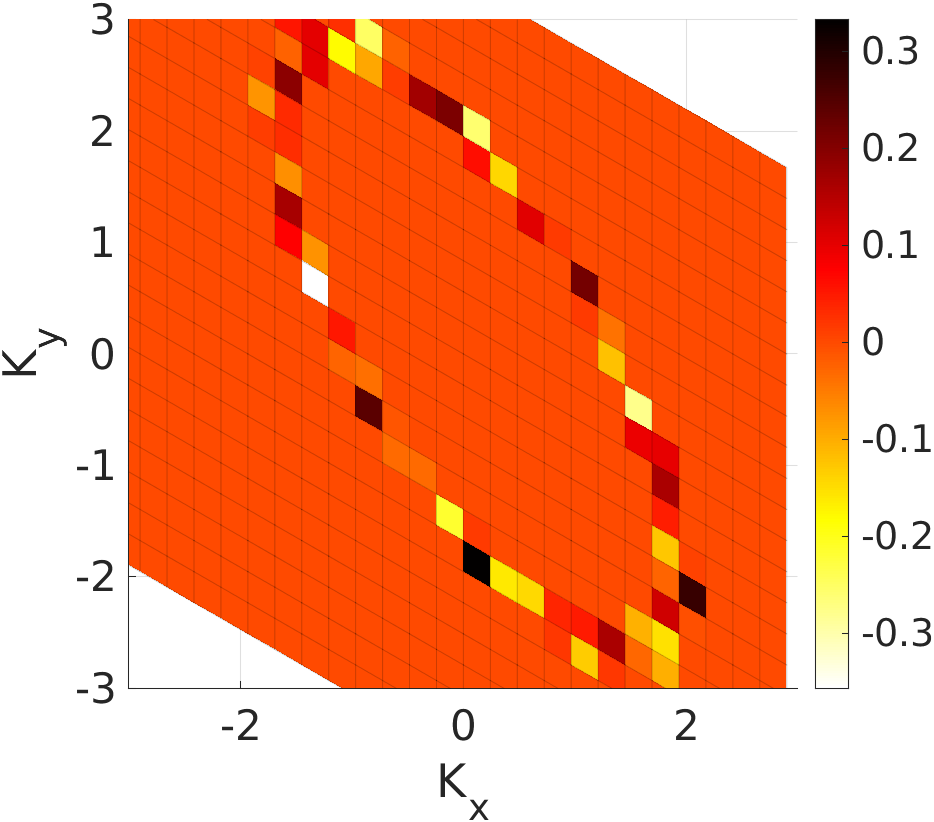}
 \label{real4f45}
 }
 \subfigure[~imag, $f=0.45$]{
  \includegraphics[scale=0.2]{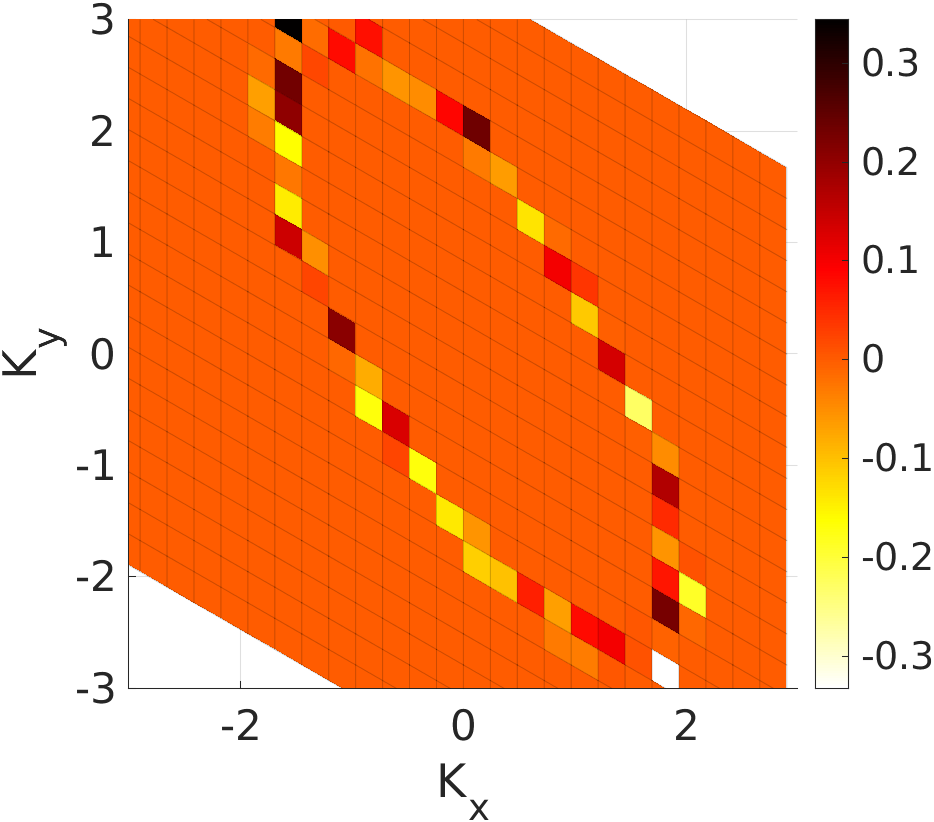}
 \label{imag4f45}
 }
 \caption{$P(k)$ condensate on the reciprocal lattice corresponding to a triangular arrangement of 
 $26\times 26$ lattice sites at coupling
 $U=4$, for peaks at $f=0.25,0.45$. Real and imaginary parts are shown. (a) Re[$P(k)$] at $f=0.25$. 
 (b) Im[$P(k)$] at $f=0.25$. (c) Re[$P(k)$] at $f=0.45$. (d) Im[$P(k)$] at $f=0.45$.}
  \label{cond4}
 \end{figure*}

\begin{figure*}[htb]
 \subfigure[~real, $f=0.25$]{
 \includegraphics[scale=0.2]{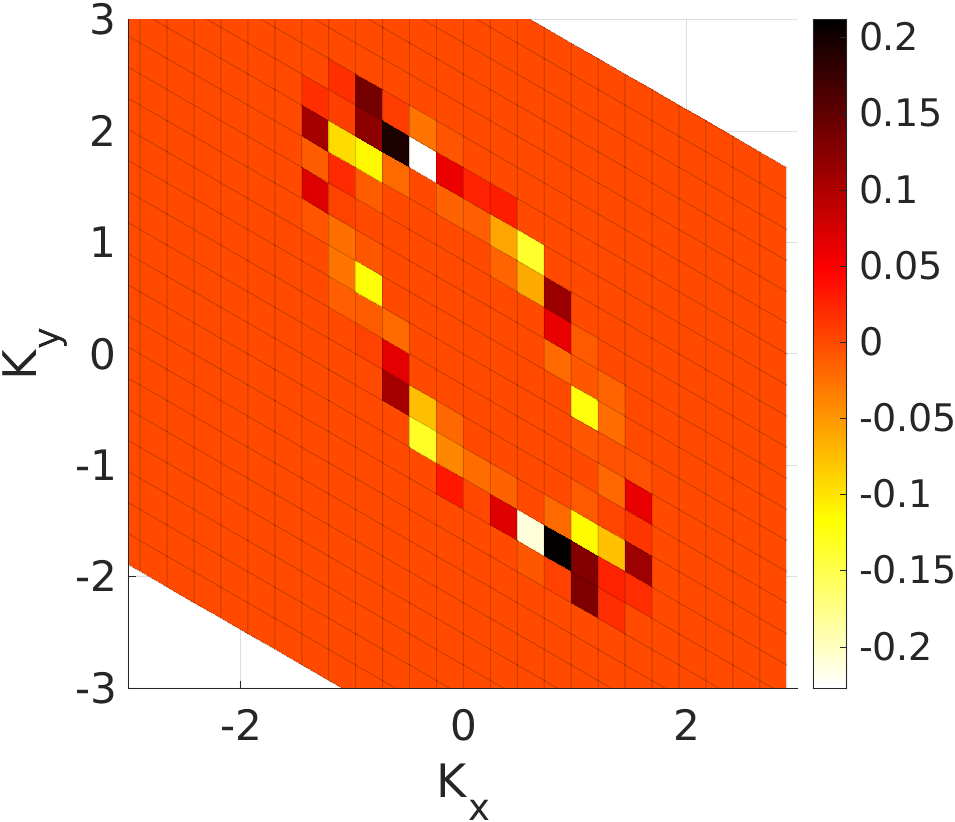}
 \label{real8f25}
 }
 \subfigure[~imag, $f=0.25$]{
   \includegraphics[scale=0.2]{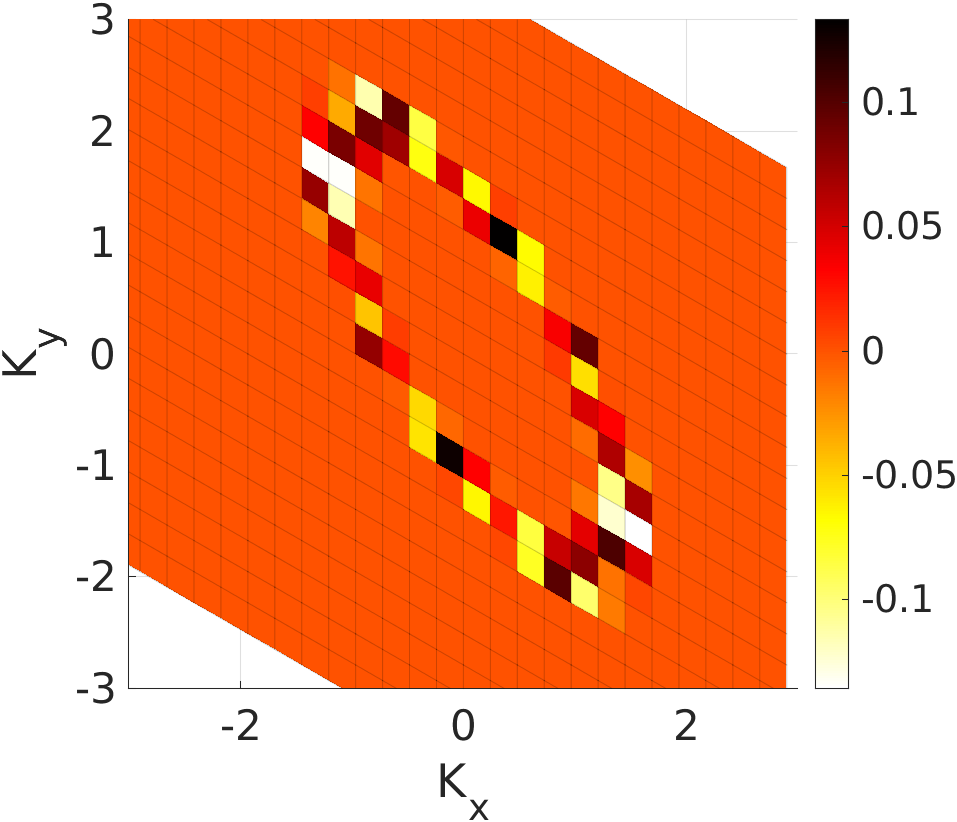}
 \label{imag8f25}
 }
  \subfigure[~real, $f=0.45$]{.
 \includegraphics[scale=0.2]{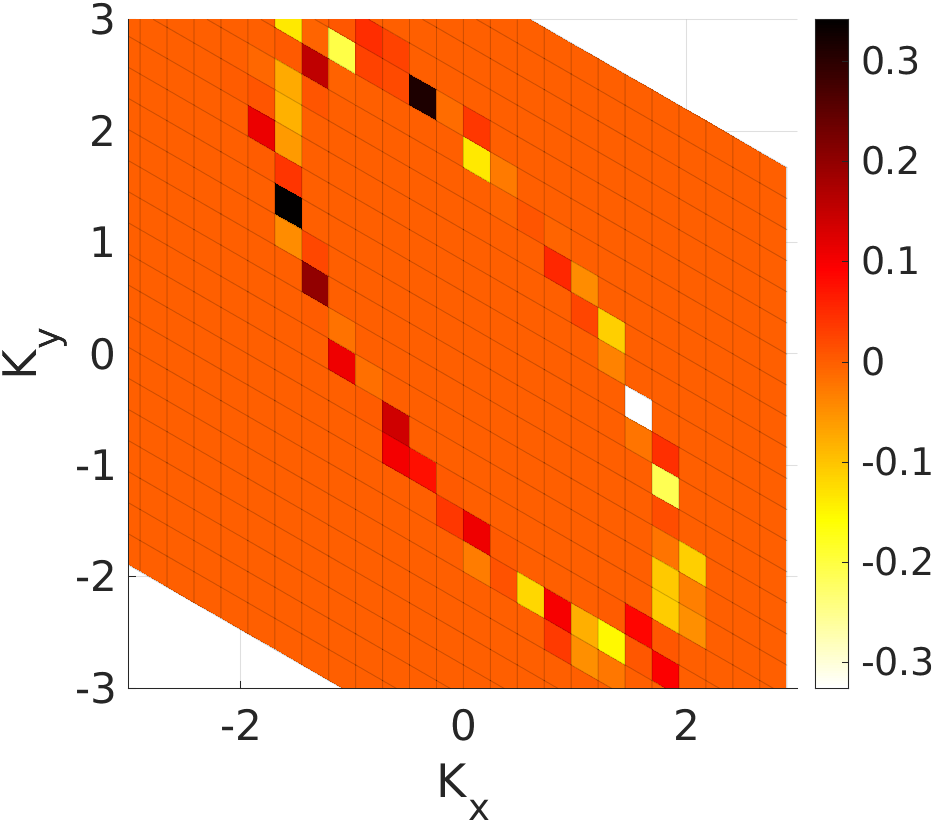}
 \label{real8f45}
 }
 \subfigure[~imag, $f=0.45$]{
  \includegraphics[scale=0.2]{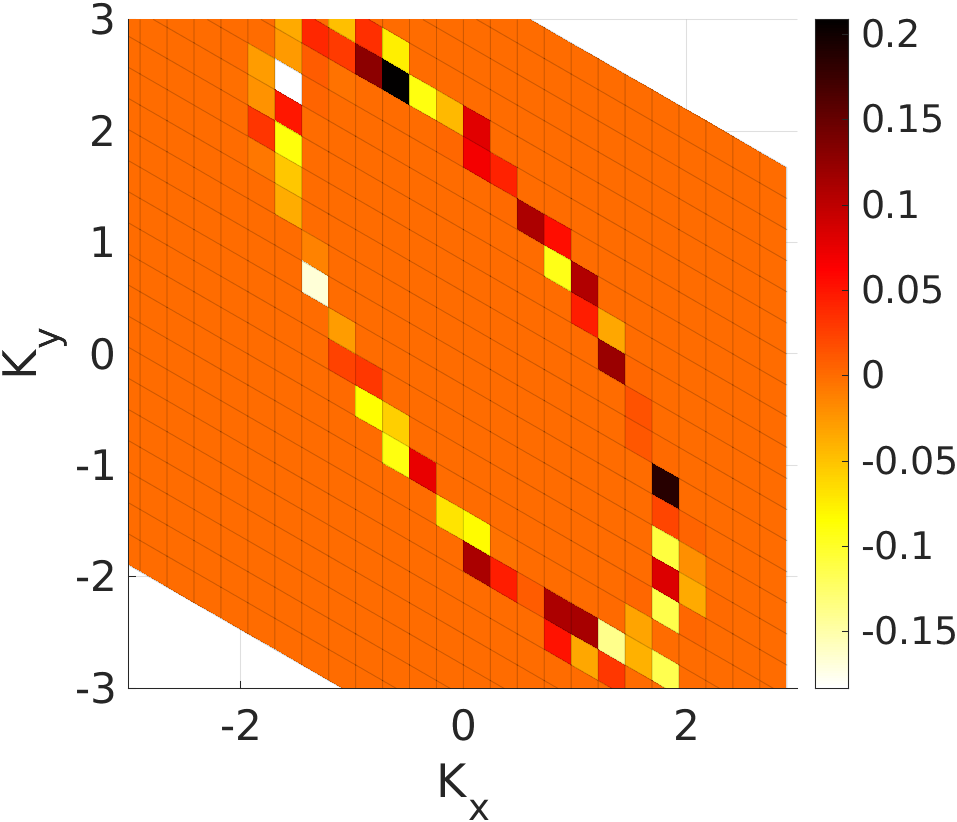}
 \label{imag8f45}
 }
 \caption{ Same as the previous figure, only at coupling $U=8$.}
  \label{cond8}
 \end{figure*}

   We have carried out computations of the condensate $P(k)$ for the triangular lattice at $U=2,4,8,12$
over a range of electron densities on a $12\times 12$ lattice (using $N=20$ for the BCS calculation) and, as a check, carried out a few computations on a $26\times 26$ ($N=80$) lattice which will be displayed below.  The pattern which results from such an apparently minor change in connectivity appears to be quite different from what is seen on the standard square lattice, with only nearest neighbor connections.  We define $P_{max}$ as the maximum value of the condensate on the reciprocal lattice, and $P_{sum}$ as sum of all values of the condensate on the $12\times 12$ lattice,
discarding values below $10^{-3}$.  Either way, the pictures are similar.  Fig.\ \ref{Pmax} shows $P_{max}$ vs.\ $f$ for $U=2,4,8,12$, and Fig.\ \ref{Psum}
is the same plot for $P_{sum}$.   What is obvious, first of all that a substantial condensate appears
for electron densities in the range $0.2 \lesssim f \lesssim 0.6$, and this holds true for all $U$ values we have looked at.  The second
obvious and striking feature is the existence of a double peak in the condensate, whose positions are only weakly dependent on $U$, at densities $f \approx 0.45$ and $f \approx 0.25$, 

     We have repeated the computation of $P(k)$ at $U=4$ and $U=8$ at these peak locations on a much
larger $26\times 26$ lattice volume.  The results in each case, for the real and imaginary parts of the condensate, are shown in Fig.\ \ref{cond4} for $U=4$, and Fig.\ \ref{cond8} for $U=8$.  Note that the reciprocal lattice is actually a periodic parallelogram.  In the displayed figures we have cut off the upper left and bottom right portions, which carry no additional information.  

 \begin{figure}[htb]
 \includegraphics[scale=0.4]{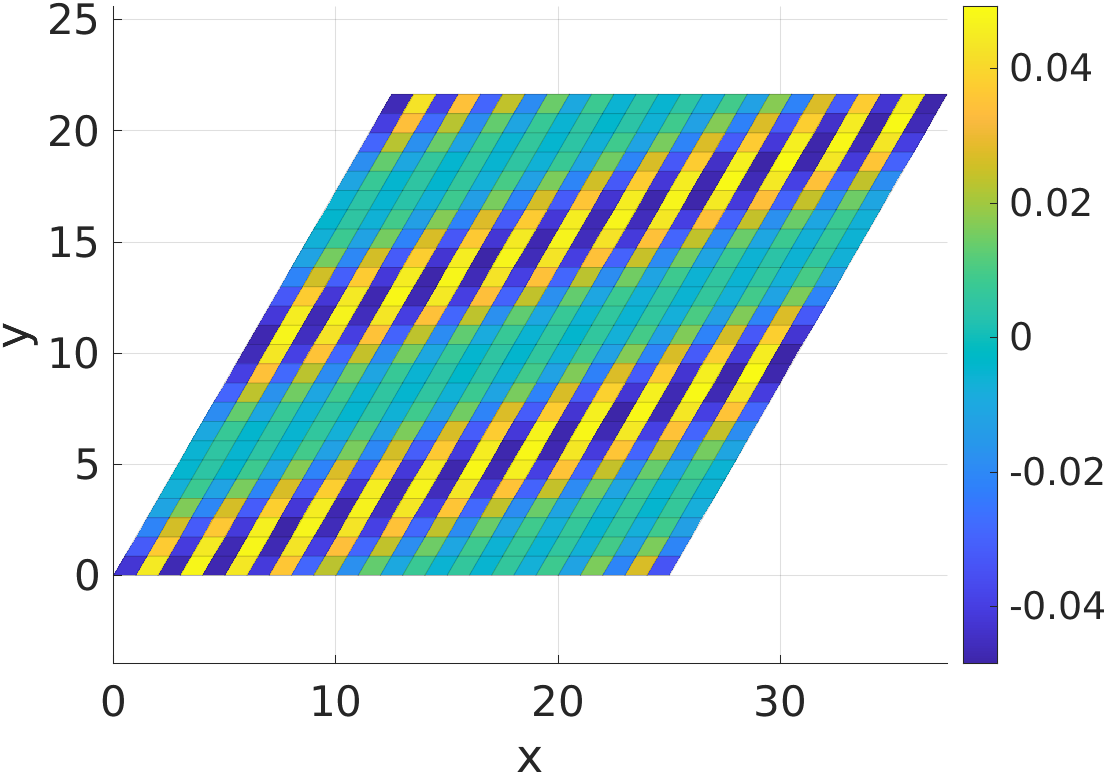}
 \caption{$D(x)$ on a triangular lattice at one of the peaks: $U=8, f=0.45$.}
 \label{trigeo}
 \end{figure}
 
  Compared to spin densities $D(x)$ shown above for the square lattice arrangement, the amplitudes of $D(x)$ on the triangular lattice is greatly reduced by at least one order of magnitude, more often by two.  An example (where we found the largest amplitude) is shown in Fig.\ \ref{trigeo}, obtained at $U=8, f=0.45$ on a $26\times 26$ lattice volume, but we emphasize again that in general the spin densities are strongly suppressed on the triangular lattice as compared to the square lattice.
 
\subsection{Limitations}

  Our Hartree-Fock BCS approach shares the limitations of ordinary Hartree-Fock, or in fact any mean field theory, namely a neglect of correlations.  We should expect that this neglect is most harmful in the strong-coupling underdoped region of the phase diagram, where a mean field approach is most likely to overestimate double occupied states, and hence significantly overestimate the energy.  This is, of course, what the Gutzwiller projection was designed to avoid.  Quantitatively we can compare the energy we compute by
our method at $U=8, f=0.875$ with the energies reported at those parameters by Zheng et al.\ \cite{Zheng} using a range of methods.  Unfortunately, but perhaps not surprisingly, the energy values obtained by our methods are greater than the reported values at this point in the phase diagram by about 22\%.  This of 
course suggests that our Hartree-Fock BCS approach might be improved in the strong coupling underdoped regime by combining it with some form of Gutzwiller projection.  We leave this possibility for future study.

\section{Discussion}

   One interesting aspect of this investigation is the appearance, on a square lattice, of a strong peak in the  d-wave condensate $\D_1$ at moderate couplings around $U/t=4$ and electron density $0.8$, and the existence of condensation in this region is consistent with some other recent results \cite{Deng,Pfaffian2}.  The peak subsides, as expected, at low density and half-filling.  The symmetry of the condensate was not built into the original BCS ansatz, and in our calculation the magnitude of the condensate subsides at large $U$.
It is also of interest that, in the overdoped region, we find evidence of $d_{xy}$ condensation in the 2D Hubbard model, both at $U=4$ and $U=8$, as opposed to the $d_{x^2-y^2}$ condensates we find in the underdoped region.  At $U=4$ we can compare with the results of Deng et al.\ \cite{Deng}, who find $d_{x^2-y^2}$ condensation in the underdoped region, and $d_{xy}$ condensation in the overdoped region, consistent with what we have found.

    We have also investigated condensation in a triangular lattice at various densities and couplings.  Here it seems that the geometry makes a drastic difference to the results, as compared to what we find on the square lattice.  Rather than being confined to a fairly small region in the neighborhood of $U=4$,
there is evidence of condensation for all $U$ values investigated ($U=2,4,8,12$) for electron densities
in the range $0.2 \lesssim f \lesssim 0.6$.  We do not understand origin of the curious double peak in the
condensate in Figs.\ \ref{Pmax} and \ref{Psum}, nor are we able to characterize the condensate seen in Figs.\ \ref{cond4} and \ref{cond8} according to any symmetry scheme we are aware of.  Conceivably the double peak is associated with some change in condensate symmetry.  One might also speculate that the condensates we are seeing are a superposition of differing symmetries, as suggested in \cite{Arovas_2022}.  For the moment we leave these questions open.
   
   We also emphasize that there are many local minima around a given Hartree-Fock state, and there are many Hartree-Fock states.  By enlarging to BCS states, what we are doing is examining the condensate properties
of states in the neighborhood of the standard local minima, which could not be easily studied in the standard states themselves.  Condensates in the Hartree-Fock state itself are not excluded by any means but are difficult to observe, since in this state, as in any eigenstate of electron number, the expectation value of $c^\dg_{s}(x) c^\dg_{s'}(y)$ and its conjugate must vanish.  Other observables  are required, e.g.\ correlators  
$\langle \D^\dg_1(x) \D_1(y)\rangle$, and the computation is more challenging, particularly on small lattices.  The main advantage of the BCS-like approach is that such non-local correlators are unnecessary for probing the existence of condensates. The multiplicity of local minima around the standard Hartree-Fock states is also relevant.  If the multiplicity of local minima in this BCS-like approximation is a reflection of a genuine multiplicity of near-degenerate energy eigenstates near the ground state, as in a spin glass, then the properties of those near-ground states may be dominant.  Whether a large multiplicity of states with energies
close to the ground state has other observable properties, as in a spin glass, is an open question.

As it stands now, each relaxation step is aimed at reducing the energy expectation of the BCS state.  It would be interesting if this relaxation criterion were revised so as to give priority towards reducing double occupancy 
at the beginning of the relaxation process, and returning to the energy criterion alone towards the end of the
procedure.  This is somewhat in the spirit of Gutzwiller projection, and we may speculate that the procedure,
which would find other paths to local minima, might attain a lower energy expectation value than the energy criterion alone, or possibly enhance the condensate at strong couplings.  Alternatively, our procedure might be followed by a Gutzwiller projection of some kind. Another line of investigation is inclusion of next-nearest couplings, with strength parametrized by $t'$.  Some studies indicate a strong dependence of condensation on this parameter \cite{Qin2}, and this dependence deserves attention in our approach, which up until now is at $t'=0$.  It would also be helpful to recompute the data leading to
Fig.\ \ref{view3d} on a larger lattice, although this computation will require a significant increase in computer time.  We leave this for future work.

\acknowledgments{This research is supported by the U.S.\ Department of Energy under Grant No.\ DE-SC0013682.}  

\bibliography{hub}

\end{document}